\documentclass[final,12pt,times]{elsarticle}

\usepackage{amssymb}
\usepackage{amsthm}
\usepackage{amsmath}
\usepackage{amscd,amsfonts,bbold}
\usepackage{caption}
\usepackage{subcaption}
\usepackage{lineno}

\journal{Journal of Computational Physics}

\begin{document}

\begin{frontmatter}



\title{Discrete exterior calculus discretization of incompressible Navier-Stokes equations over surface simplicial meshes}


\author[address1]{Mamdouh S. Mohamed\corref{cor1}\fnref{fn01}}
\ead{mamdouh.mohamed@kaust.edu.sa}
\author[address2]{Anil N. Hirani}
\ead{hirani@illinois.edu}
\author[address1]{Ravi Samtaney}
\ead{ravi.samtaney@kaust.edu.sa}

\address[address1]{Mechanical Engineering, Physical Sciences and Engineering Division, KAUST, Jeddah, KSA}
\address[address2]{Department of Mathematics, University of Illinois at Urbana-Champaign, IL, USA}

\cortext[cor1]{Corresponding author}
\fntext[fn01]{On leave, Department of Mechanical Design and Production, Faculty of Engineering, Cairo
University, Giza, Egypt.}

\begin{abstract}
A conservative discretization of incompressible Navier-Stokes equations is developed based on discrete exterior calculus (DEC). A distinguishing feature of our method is the use of an algebraic discretization of the interior product operator and a combinatorial discretization of the wedge product. The governing equations are first rewritten using the exterior calculus notation, replacing vector calculus differential operators by the exterior derivative, Hodge star and wedge product operators. The discretization is then carried out by substituting with the corresponding discrete operators based on the DEC framework. Numerical experiments for flows over surfaces reveal a second order accuracy for the developed scheme when using structured-triangular meshes, and first order accuracy for otherwise unstructured meshes. By construction, the method is conservative in that both mass and vorticity are conserved up to machine precision. The relative error in kinetic energy for inviscid flow test cases converges in a second order fashion with both the mesh size and the time step.  
\end{abstract}


\end{frontmatter}


\section{Introduction}
\label{sec:introduction}

When solving a partial differential equation numerically, various measures (e.g. convergence, stability and consistency) are usually investigated to verify the implemented discretization. Such measures, although reflecting the mathematical fidelity of the discretization, may not give insight into the physical fidelity of the discretization. By physical fidelity we mean how well does the discrete system of equations conserve secondary quantities, such as kinetic energy, implied in the continuous equation but not explicitly constructed or built into the numerical scheme. The development of such physically conservative discretizations, for Navier-Stokes (NS) equations for example, is favorable for many physical applications (e.g. turbulent flows and shallow-water simulations) to avoid undesirable numerical artifacts. Among other discretization approaches, some staggered mesh schemes are known for their conservation of both primary (i.e. mass and momentum) and secondary (e.g. kinetic energy and vorticity) physical quantities \cite{Perot:2011dcpums}.  

Staggered mesh methods were first developed by Harlow and Welch \cite{HarlowWelch:1965} for structured Cartesian meshes by placing the velocity and pressure degrees of freedom at different positions on the mesh. Later on, the approach was extended to unstructured meshes by Nicolaides \cite{nicolaides:1989flow} and Hall \textit{et al.} \cite{HallCavendish:1991dvmsvf}, which is now known as the covolume (or dual-variable) method. The covolume method was originally introduced as a low order method that does not experience spurious modes that were common in low order discretizations of viscous flows. The derivation of such discretization commences by taking the dot product of the momentum equation with the unit normal vector to each face of the triangular/tetrahedral elements. This reduces the velocity vector field to a scalar flux (equal to the mass flux across the face for an incompressible flow with constant density) defined on each face. In this approach, the pressure is consequently defined at the circumcenter of each triangular/tetrahedral element. The accuracy of the covolume scheme was estimated by Nicolaides \cite{nicolaides:1992direct} to be second order for a mesh with all its triangular elements having the same circumradii (i.e. structured-triangular mesh) and first order accurate otherwise. These accuracy estimates were in agreement with numerical experiments \cite{HallCavendish:1991dvmsvf}. Several forms of the covolume method were then developed for both two-dimensional (2D) (only on planar domains) and three-dimensional (3D) domains, where the difference was mainly in the convective term discretization \cite{cavendish1992solution,cavendish1994complementary,choudhury1990discretization,nicolaides1993covolume}. 

The conservation properties of the covolume method were later revealed by Perot \cite{perot2000conservation}. The divergence form of Navier-Stokes equations was proved to conserve the momentum and kinetic energy both locally and globally. On the other hand, the rotational form of Navier-Stokes equations was found to conserve the circulation and kinetic energy locally and globally for both 2D \cite{perot2000conservation} and 3D \cite{zhang2002accuracy} discretizations. These conservation properties of the covolume method, in addition to the attractive properties of its differential operators that mimic the behavior of their continuous counterparts, shed light on the merit of using discrete calculus methods to solve other physics problems \cite{perot2007discrete}.         

Another approach to develop conservative discretizations for incompressible flows emerged from the computer graphics community, aiming to mitigate the effects of numerical viscosity that causes detrimental visual consequences \cite{elcott2007stable,mullen2009energy}. In this approach, the Navier-stokes equations were discretized through the discrete exterior calculus (DEC) framework; the discretization of the smooth exterior calculus operators \cite{hirani2003discrete,desbrun2005discrete}. A main advantage of DEC discretizations is the applicability to simulate flows over curved surfaces, unlike the covolume approach. The resulting discrete equations had similarities with the covolume method, with the differences mainly in the convective term discretization. In practice, the convective term was not discretized using DEC but employed a method of characteristics with an interpolation scheme based on Kelvin's circulation theorem \cite{elcott2007stable}, or a finite volume based approach \cite{mullen2009energy}. However, the presented numerical test cases using the DEC approach lacked comprehensive quantitative analysis of the scheme's accuracy and its conservative behavior.

This paper presents a discretization of the Navier-Stokes equations through discrete exterior calculus. Hence, similar to previous DEC-based discretizations, the developed discretization is capable of simulating flows over curved surfaces, which distinguishes it from the covolume method. In addition, the convective term in the presented discretization is different from previous DEC-based and covolume discretizations. The Navier-Stokes equations are first rephrased using the exterior calculus notation in Section \ref{sec:smoothNS}. The DEC discretization of NS equations is then derived in Section \ref{sec:discretization} for both 2D and 3D cases, highlighting its distinction from the covolume method and previous DEC-based discretizations. In Section \ref{sec:results}, several numerical experiments are illustrated for incompressible flows over 2D flat/curved domains to benchmark the convergence and conservative behavior of the developed scheme. The paper closes with conclusions summarizing the main features of the presented discretization, and addressing potential future developments.

\section{Navier-Stokes equations in exterior calculus notation}
\label{sec:smoothNS}

The first step in deriving a DEC discretization of NS equations is to express the equations using the exterior calculus notation. This is done first by starting from the well-known vector calculus formulation of NS equations (in Euclidean space) and substituting with identities relating the differential operators; viz. div, grad and curl, with exterior calculus operators; viz. exterior derivative, Hodge star and wedge product. An alternative derivation of the resultant formulation is then presented, starting from the coordinate invariant formulation of NS equations expressed in terms of the Lie and exterior derivatives. Readers not familiar with exterior calculus may refer to \cite{HiNaCh2015,perot2014differential} for a concise introduction to the topic.   
 
Considering the incompressible flow of a homogeneous fluid with unit density and no body forces, the governing equations for the fluid motion are the Navier-Stokes equations expressed as

\begin{subequations}
\label{eq:NS-vector-calculus-1}
\begin{align}
\label{eq:NS-vector-calculus-1-a}
&\frac{\partial \mathbf{u}}{\partial t} - \mu \Delta \mathbf{u} + (\mathbf{u}.\nabla) \mathbf{u} + \nabla p =0, \\
\label{eq:NS-vector-calculus-1-b}
&\nabla . \mathbf{u} = 0,
\end{align}
\end{subequations}
where $\mathbf{u}$ is the velocity vector, $p$ is the pressure and $\mu$ is the dynamic viscosity. Substituting with the tensor identities: $\Delta \mathbf{u} = \nabla (\nabla . \mathbf{u}) - \nabla \times (\nabla \times \mathbf{u})$ and $(\mathbf{u}.\nabla) \mathbf{u} = \frac{1}{2}\nabla( \mathbf{u}.\mathbf{u}) - \mathbf{u} \times (\nabla \times \mathbf{u})$, and considering the incompressibility condition $\nabla . \mathbf{u} = 0$, Eq. \eqref{eq:NS-vector-calculus-1-a} can be expressed in its rotational form as

\begin{equation}
\label{eq:NS-vector-calculus-2}
\frac{\partial \mathbf{u}}{\partial t} + \mu \nabla \times \nabla \times \mathbf{u} - \mathbf{u} \times (\nabla \times \mathbf{u}) + \nabla p^d =0,
\end{equation}
where $p^d$ is the dynamic pressure defined as $p^d = p + \frac{1}{2}( \mathbf{u}.\mathbf{u})$. 

The notation transformation is carried out by applying the flat operator $(\flat)$ to Eqs. \eqref{eq:NS-vector-calculus-2} and \eqref{eq:NS-vector-calculus-1-b}, and substituting with the following identities

\begin{equation}
\begin{aligned}
\label{eq:vec-cal-ext-cal}
(\nabla \times \nabla \times \mathbf{u})^\flat &= (-1)^{N+1}  \ast \textrm{d} \ast \textrm{d} \mathbf{u}^\flat,  \\
(\mathbf{u} \times (\nabla \times \mathbf{u}))^\flat &= (-1)^{N+1}  \ast(\mathbf{u}^\flat \wedge \ast \textrm{d} \mathbf{u}^\flat),\\
(\nabla . \mathbf{u})^\flat &= \ast \textrm{d} \ast \mathbf{u}^\flat, \\
(\nabla p^d)^\flat &= \textrm{d}p^d, 
\end{aligned} 
\end{equation} 
where $\ast$ is the Hodge star, $\textrm{d}$ is the exterior derivative, and $\wedge$ is the wedge product operators. The action of the flat operator $(\flat)$ on a vector $\mathbf{u}$ transforms it into a 1-form $\mathbf{u}^\flat$. The sign $(-1)^{N+1}$ in the first two relationships implies a negative sign only in the two-dimensional (2D) case, where $N$ is the space dimension (i.e. $N$ = 2 and 3 for the 2D and 3D cases, respectively). The above relationships can be easily verified using the definition of differential forms and the action of exterior derivative, Hodge star and wedge product operators on them. Substituting with Eqs. \eqref{eq:vec-cal-ext-cal} in Eqs. \eqref{eq:NS-vector-calculus-2} and \eqref{eq:NS-vector-calculus-1-b}, the Navier-Stokes equations are then expressed as

\begin{subequations}
\label{eq:NS_DG-00}
\begin{align}
\label{eq:NS_DG-00-a}
&\frac{\partial \mathbf{u}^\flat}{\partial t} + (-1)^{N+1} \mu \ast \textrm{d} \ast \textrm{d} \mathbf{u}^\flat + (-1)^{N+2} \ast(\mathbf{u}^\flat \wedge \ast \textrm{d} \mathbf{u}^\flat) + \textrm{d}p^d =0,   \\
\label{eq:NS_DG-00-b}
&\ast \textrm{d} \ast \mathbf{u}^\flat = 0,
\end{align}
\end{subequations}
where the velocity field is represented by the 1-form $\mathbf{u}^\flat$ and $p^d$ is now the dynamic pressure 0-form.  

The above formulation in Eqs. \eqref{eq:NS_DG-00} is derived starting from the standard vector calculus formulation of NS equations. We now present an alternative derivation purely in terms of differential forms, starting from NS equations formulated using the exterior derivative ($\textrm{d}$) and the Lie derivative ($\pounds$) operators. The Navier-Stokes equation in coordinate invariant form is (see \cite{AbMaRa1988} pg. 589 for Euler equation in this form)

\begin{equation}
\label{eq:NS_DG01}
\frac{\partial \mathbf{u}^\flat}{\partial t} + \mu (\delta \textrm{d}+\textrm{d} \delta) \mathbf{u}^\flat + \pounds_{\mathbf{u}} \mathbf{u}^\flat - \frac{1}{2} \textrm{d} (\mathbf{u}^\flat(\mathbf{u}))  + \textrm{d}p =0,  
\end{equation}    
where $\delta$ is the codifferential operator defined as $\delta = (-1)^{N(k-1)+1} \ast \textrm{d} \ast$, which acts on a $k$-form and results in a $(k-1)$-form, and $N$ again is the space dimension. Therefore, the incompressibility condition (Eq. \eqref{eq:NS_DG-00-b}) translates to $\delta \mathbf{u}^\flat = 0$. The operator ($\delta \textrm{d}+\textrm{d} \delta$) is the Laplace operator in the exterior calculus notation, which differs by a negative sign from that defined in the vector calculus notation (e.g. the Laplace operator in Eq. \eqref{eq:NS-vector-calculus-1-a}). The Lie derivative term ($\pounds_{\mathbf{u}} \mathbf{u}^\flat$) evaluates the change of the velocity 1-form $\mathbf{u}^\flat$ along the velocity vector field $\mathbf{u}$, and the term $\mathbf{u}^\flat(\mathbf{u})$ represents the dot product of the vector field $\mathbf{u}$ with itself. Using Cartan homotopy formula (see \cite{AbMaRa1988}, pg. 430), the Lie derivative term is expressed as 

\begin{equation}
\label{eq:CartanHomotopy}
\pounds_{\mathbf{u}} \mathbf{u}^\flat = \textrm{d} i_{\mathbf{u}} \mathbf{u}^\flat + i_{\mathbf{u}} \textrm{d} \mathbf{u}^\flat = \textrm{d}(\mathbf{u}^\flat(\mathbf{u})) + i_{\mathbf{u}} \textrm{d} \mathbf{u}^\flat,  
\end{equation}
where $i_{\mathbf{x}} \alpha$ is the interior product of a $k$-form $\alpha$ with a vector field $\mathbf{x}$. Accordingly, considering the incompressibility condition $\delta \mathbf{u}^\flat = 0$, Eq. \eqref{eq:NS_DG01} can be expressed as

\begin{equation}
\label{eq:NS_DG02}
\frac{\partial \mathbf{u}^\flat}{\partial t} + \mu \delta \textrm{d} \mathbf{u}^\flat + i_{\mathbf{u}} \textrm{d} \mathbf{u}^\flat + \frac{1}{2} \textrm{d} (\mathbf{u}^\flat(\mathbf{u}))  + \textrm{d}p =0.  
\end{equation}

Defining the dynamic pressure 0-form as $p^d = p + \frac{1}{2} (\mathbf{u}^\flat(\mathbf{u}))$, and substituting with $\delta = (-1)^{N+1} \ast \textrm{d} \ast$ since the codifferential operator $\delta$ in Eq. \eqref{eq:NS_DG02} acts on the 2-form $\textrm{d} \mathbf{u}^\flat$ (hence $k=2$), Eq. \eqref{eq:NS_DG02} then takes the form

\begin{equation}
\label{eq:NS_DG03}
\frac{\partial \mathbf{u}^\flat}{\partial t} + (-1)^{N+1} \mu \ast \textrm{d} \ast \textrm{d} \mathbf{u}^\flat + i_{\mathbf{u}} \textrm{d} \mathbf{u}^\flat  + \textrm{d}p^d =0.  
\end{equation} 

The interior product term $i_{\mathbf{u}} \textrm{d} \mathbf{u}^\flat$ can be written in terms of the Hodge star and wedge product using the formula (see \cite{hirani2003discrete} Eq. 8.2.1)
\begin{equation}
\label{eq:interior-product}
i_{\mathbf{x}} \alpha = (-1)^{k(N-k)} \ast (\ast \alpha \wedge \mathbf{x}^\flat),  
\end{equation} 
for a $k$-form $\alpha$ and a vector field $\mathbf{x}$. Therefore, with $\mathbf{x} = \mathbf{u}$ and $\alpha = \textrm{d} \mathbf{u}^\flat$ (hence $k=2$), Eq. \eqref{eq:NS_DG03} becomes

\begin{equation}
\label{eq:NS_DG04}
\frac{\partial \mathbf{u}^\flat}{\partial t} + (-1)^{N+1} \mu \ast \textrm{d} \ast \textrm{d} \mathbf{u}^\flat + (-1)^{N-2} \ast(\mathbf{u}^\flat \wedge \ast \textrm{d} \mathbf{u}^\flat)  + \textrm{d}p^d =0,  
\end{equation} 
where the order of the wedge product was flipped using the relation $\alpha \wedge \beta = (-1)^{kl} \beta \wedge \alpha$, with $\alpha = \ast \textrm{d} \mathbf{u}^\flat$ is an $(N-2)$-form and $\beta = \mathbf{u}^\flat$ is a 1-form. Noting that $(-1)^{N+2} = (-1)^{N-2} = (-1)^{N}$ for any $N$, Eq. \eqref{eq:NS_DG04} is exactly the same as Eq. \eqref{eq:NS_DG-00-a}, which was derived earlier starting from the standard vector calculus formulation of Navier-Stokes equation.

Applying the exterior derivative operator to Eq. \eqref{eq:NS_DG04} (equivalent to taking the curl of the momentum equation Eq. \eqref{eq:NS-vector-calculus-2}), and considering the exterior derivative property $\textrm{d}\textrm{d}=0$, the resulting governing equation is

\begin{equation}
\label{eq:vorticityNS}
\frac{\partial \textrm{d} \mathbf{u}^\flat}{\partial t} + (-1)^{N+1} \mu \textrm{d} \ast \textrm{d} \ast \textrm{d} \mathbf{u}^\flat + (-1)^{N} \textrm{d} \ast(\mathbf{u}^\flat \wedge \ast \textrm{d} \mathbf{u}^\flat) =0. 
\end{equation}  
The DEC discretization of Navier-Stokes equations is carried out through the discretization of Eq. \eqref{eq:vorticityNS}. The advantage of discretizing the vorticity form of the NS equations is highlighted in the next section.
\section{The discretization method}
\label{sec:discretization}

In this section, the notation of the simplicial mesh used to discretize the simulation domain is first introduced. This is followed by brief description of the discrete differential forms and some of the discrete operators. The discretization of NS equations is then derived for both 2D and 3D cases. The simplicial mesh and discrete operators concepts are discussed only briefly here. Readers interested in more details may refer to \cite{hirani2003discrete,desbrun2005discrete,HiNaCh2015}.     

\subsection{The domain discretization}
\label{subsec:domain-discretization}

\begin{figure}
\centering
\includegraphics[scale=0.12]{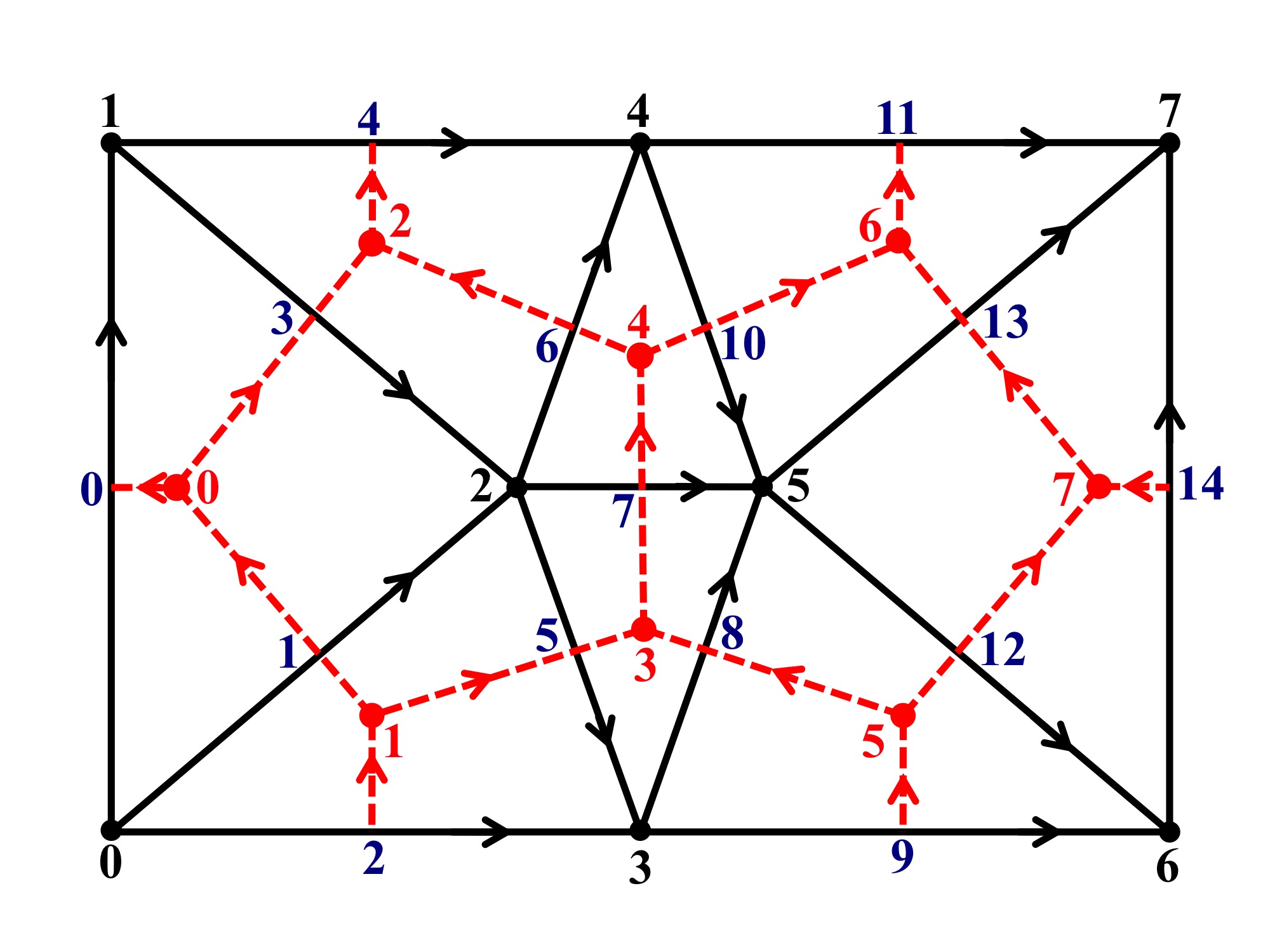}
\caption{A sample simplicial mesh in 2D showing the primal simplices (in black color) and their dual cells (in red color). The positive orientation of the primal $2$-simplices and dual $2$-cells is counterclockwise.}
\label{fig:mesh2D}
\end{figure}

Let $\Omega$ be the physical domain of dimension $N=2$ or $3$. The domain $\Omega$ is approximated by the simplicial complex $K$. Following the notation in \cite{hirani2003discrete,HiNaCh2015}, a domain simplex $\sigma$ of dimension $k$ is denoted by $\sigma^k \in K$. A $k$-simplex $\sigma^k$ is defined by the nodes forming it as $\sigma^k = [v_0, ..., v_k]$, where the subscripts represent the nodes indices. The order of the nodes defining a simplex implies its orientation. The top dimensional simplices $\sigma^N$ are assumed to have been oriented consistently, whereas the orientation of the lower dimensional simplices is arbitrary. An example 2D mesh is shown in Fig. \ref{fig:mesh2D}. The number of $k$-simplices in the discrete mesh is denoted by $N_k$. Therefore for the mesh in Fig. \ref{fig:mesh2D}, $N_2 = 8, N_1 = 15,$ and $N_0 = 8$.     

Associated with the primal simplicial complex $K$ is a dual complex $\star K$ consisting of cells. For a primal $k$-simplex $\sigma^k \in K$, its dual is an $(N-k)$-cell denoted by $\star \sigma^k \in  \star K$. The dual mesh considered here is the circumcentric dual, shown in Fig. \ref{fig:mesh2D} in red color. For a 2D mesh, the dual of a triangle is its circumcenter, the dual of a primal edge is the dual edge connecting the circumcenters of the two triangles sharing the primal edge, and the dual of a primal node is the 2-cell (polygon) formed by the duals of the edges connected to this primal node. For the case of a triangular mesh representing a curved surface, the dual edges can be kinked lines and the dual cells can be non planar. In 2D, the positive orientation of both primal $2$-simplices (triangles) and dual $2$-cells (polygons) is assumed to be counterclockwise. The orientation of the primal 1-simplices (edges) is arbitrary, however their orientations induce the dual edges orientations. The dual edges can be oriented simply by rotating the primal edge orientation 90 degrees along the triangles orientation (i.e. counterclockwise), as shown in Fig. \ref{fig:mesh2D}. In the 3D case, the orientations of the primal edges and faces (i.e. triangles) are arbitrary, however, their orientations induce the orientations of their duals through the right hand rule.   

The simplicial meshes considered here are Delaunay meshes, with an extra requirement only for the $N$-simplices with a face on $\partial \Omega$, the domain boundary. Previous investigation \cite{hirani2013delaunay} showed that in order to correctly represent the discrete Hodge star operator, the mesh interior $N$-simplices have to be pairwise Delaunay, while the $N$-simplicial elements with a face on the domain boundary have to be one-sided (i.e., with respect to the face of the $N$-simplex on the domain boundary, both the simplex circumcenter and its apex have to be on the same side). Such Delaunay meshes can easily be generated using commercial or open source mesh generators. 

\subsection{Discrete exterior calculus}
\label{subsec:DEC}

Discrete exterior calculus provides discrete definitions to many of the exterior calculus operators (e.g., exterior derivative, Hodge star, wedge product, etc.) \cite{hirani2003discrete,desbrun2005discrete}. These discrete operators have the advantage that they satisfy many of the rules/identities that characterizes their smooth counterparts. Such mimetic behavior of the discrete operators is known to result in preserving the physics implied in the smooth governing equations at the discrete level \cite{perot2007discrete}, which consequently improves the physical fidelity of the numerical discretization method.

The key entities in exterior calculus are differential forms, which according to Flanders \cite{flanders2012differential} are best thought of as ``the things which occur under integral signs" \cite{perot2014differential}. For example, considering the integration of a scalar function over a three dimensional space; i.e. $\int_{\Omega} a \ dV$, an example of a 3-form is $b^3 = a \ dV$, where the superscript $3$ indicates a 3-form. Similar examples for 2-forms/1-forms can be deduced from integration of a vector field over areas/lines, whereas a 0-form represents a scalar function. While a smooth differential form can be integrated on a $k$-dimensional domain, the evaluation of a discrete $k$-form can be thought of as the integration carried out on discrete $k$-dimensional mesh objects; i.e., line, area or volume. Therefore, a discrete differential $k$-form ultimately associates a scalar with a discrete $k$-dimensional mesh object through integration. For example, for the smooth velocity 1-form $\mathbf{u}^\flat$, its discretization can be defined on primal edges $\sigma^1$ as $\left[ \int_{\sigma^1} \mathbf{u} . d \mathbf{l} \right]$, or on dual edges $\star \sigma^1$ as $\left[ \int_{\star \sigma^1} \mathbf{u} . d \mathbf{l} \right]$, representing a primal or dual discrete 1-form, respectively. Similarly, discrete 0-forms are defined as scalars on the primal or dual nodes, and the discrete 2-forms are scalars resulting from the integration of smooth 2-forms on primal 2-simplices (triangles) or dual 2-cells (polygons).

The space of discrete $k$-forms defined on primal and dual mesh complexes is denoted by $C^k(K)$ and $D^k(\star K)$, respectively. Such spaces are related via the discrete exterior derivative and Hodge star operators as shown in the following diagrams for both 2D and 3D cases:

\begin{equation}
\label{eq:cochncmplx2d}
\begin{CD}
C^0(K) @> \textrm{d}_0>> C^1(K) @> \textrm{d}_1>> C^2(K) \\
@VV\ast_0 V    @VV\ast_1 V    @VV\ast_2 V \\
D^2(\star K) @<-\textrm{d}_0^T< < D^1(\star K) @<\textrm{d}_1^T< <
D^0(\star K)
\end{CD}
\end{equation}

\begin{equation}
\label{eq:cochncmplx3d}
\begin{CD}
C^0(K) @>\textrm{d}_0>> C^1(K) @>\textrm{d}_1>> C^2(K) @>\textrm{d}_2>> C^3(K)\\
@VV\ast_0 V    @VV\ast_1 V    @VV\ast_2 V @VV\ast_3 V \\
D^3(\star K) @<\textrm{d}_0^T< < D^2(\star K) @<\textrm{d}_1^T< <
D^1(\star K) @<\textrm{d}_2^T< < D^0(\star K)
\end{CD}
\end{equation}

The discrete exterior derivative operator $\textrm{d}_k$ maps primal $k$-forms to primal $(k+1)$-forms. The discrete exterior derivative operator that maps dual $k$-forms to dual $(k+1)$-forms is the transpose of the $\textrm{d}_{(N-k-1)}$ operator, with a negative sign only for the $\textrm{d}_0^T$ operator in 2D (due to the defined mesh orientation convention). The discrete Hodge star operator $\ast_k$ maps primal $k$-forms to dual $(N-k)$-forms. The inverse map of the discrete $\ast_k$ operator is $\ast^{-1}_k$, which maps dual $(N-k)$-forms to primal $k$-forms. 

The discrete exterior derivative operator $\textrm{d}_k$ is a sparse matrix that is defined as the transpose boundary operator for the $(k+1)$-simplices. For example, for the 2D mesh in Fig \ref{fig:mesh2D}, the discrete $\textrm{d}_1$ operator, that maps the primal 1-forms defined on the primal edges to 2-forms defined on the triangles, is an $N_2 \times N_1$ matrix defined as

\begin{equation}
\label{eq:d1-matrix}
\begin{aligned}
\left[ \textrm{d}_1 \right]_{ij} &=\begin{cases}
    +1, & \text{if the edge $j$ is a face to the triangle $i$,} \\ &\text{and their orientations are consistent},\\
    -1, & \text{if the edge $j$ is a face to the triangle $i$,} \\ &\text{and their orientations are not consistent}, \\
     0, & \text{if the edge $j$ is not a face to the triangle $i$}.
  \end{cases}   
\end{aligned}
\end{equation}
For the primal nodes laying on the domain boundary; e.g. node 3 in Fig. \ref{fig:mesh2D}, the boundary of their dual 2-cells (polygons) includes primal boundary edges. Accordingly, the $[-\textrm{d}_0^T]$ matrix, represented by the transpose boundary operator of these dual 2-cells, is complemented by an additional operator accounting for the primal boundary edges. The discrete Hodge star operator $\ast_k$, on the other hand, is a diagonal matrix with the i-th diagonal element being the ratio between the volume of the dual $(N-k)$-simplex $\star \sigma^k_i$ and the volume of its primal $k$-simplex $\sigma^k_i$; i.e. $\frac{|\star \sigma^k_i|}{|\sigma^k_i|}$. In regards to the wedge product operator, its discrete definition is provided within the discretization presentation in the next two subsections.

\subsection{Two dimensional discretization}
\label{subsec:discretization:2D}

The two-dimensional DEC discretization of Navier-Stokes equations is derived in this subsection and the three-dimensional discretization in the next subsection. As pointed out by Hirani \textit{et al.} \cite{HiNaCh2015}, due to the intrinsic coordinate independent nature of exterior calculus, the derivation below, for the 2D case, results in a numerical method that works without change for both planar domains and curved surfaces. The dimension of the embedding space does not matter, neither do the details of the embedding. This characterizes a key distinction between the DEC-based approach and the covolume method. In the latter, the discretization of NS equations was commenced by taking the dot product of the momentum equation and the vectors perpendicular to the triangles' faces. Such normal vectors may not be unique for a simplicial mesh approximation of a curved surface.  

The discretization of NS equations is carried out here following the exact fractional step method \cite{HallCavendish:1991dvmsvf,chang2002analysis}. This consists mainly of two steps, the first is to discretize the vorticity formulation of NS equations (Eq. \eqref{eq:vorticityNS}), and the second is to substitute the velocity by its definition as the curl of a stream function, where the latter in its discrete manifestation becomes the unknown degrees of freedom in the resulting linear system. Sufficient details are provided  since the numerical experiments presented in this paper are limited only to two-dimensional flow test cases.  

Discrete exterior calculus discretization of Eq. \eqref{eq:vorticityNS} first requires the location on the mesh where one defines the discrete variables such that the smooth forms are satisfied in an integrated sense. Then, the smooth forms are replaced with their discrete counterparts, and the smooth operators are substituted by the appropriate discrete operators. Starting with the time derivative term in Eq. \eqref{eq:vorticityNS}, we choose to define the 2-form $\textrm{d} \mathbf{u}^\flat$ on the dual 2-cells. Accordingly, all the other terms of Eq. \eqref{eq:vorticityNS} are constrained to be defined on the dual 2-cells for consistency. Back to the time derivative term, it then follows, according to the diagram in Eq. \eqref{eq:cochncmplx2d}, that the velocity 1-form $\mathbf{u}^\flat$ in this term is defined on the dual 1-cells (i.e. dual edges). We denote the velocity 1-form $\mathbf{u}^\flat$ defined on the dual edges by $u$, which represents the integration of the velocity vector field along the dual edges; i.e. $u = \int_{\star \sigma^1} \mathbf{u} . d \mathbf{l} \in D^1(\star K)$.  The velocity 1-form $u$ may be referred to as the normal velocity form or the flux, since it represents the velocity normal to the triangles' faces (i.e. primal edges). Similarly for the viscous term in Eq. \eqref{eq:vorticityNS}, it follows from the diagram in Eq. \eqref{eq:cochncmplx2d} that the velocity 1-form in this term is also defined on the dual edges. 

In regards to the convective term, because the term $\textrm{d} \ast(\mathbf{u}^\flat \wedge \ast \textrm{d} \mathbf{u}^\flat)$  is defined on the dual 2-cells, the term $\ast(\mathbf{u}^\flat \wedge \ast \textrm{d} \mathbf{u}^\flat)$ has to be defined on the dual edges, and therefore $(\mathbf{u}^\flat \wedge \ast \textrm{d} \mathbf{u}^\flat)$ is defined on the primal edges. Since $\mathbf{u}^\flat$ is a 1-form, then $\ast \textrm{d} \mathbf{u}^\flat$ is a 0-form (in 2D), and therefore the wedge product is carried out between a 1-form and a 0-form, such that the outcome of this wedge product must be defined on the primal edges. This implies that both wedge product portions are defined on primal mesh objects. Accordingly, the velocity 1-form in the first portion of the wedge product term is defined on the primal edges, whereas the velocity 1-form in the second portion of the wedge product term is defined on the dual edges, making the 0-form $\ast \textrm{d} \mathbf{u}^\flat$ to be defined on the primal nodes. We denote the velocity 1-form $\mathbf{u}^\flat$ defined on the primal edges by $v$; i.e. $v = \int_{\sigma^1} \mathbf{u} . d \mathbf{l} \in C^1(K)$. The velocity 1-form $v$ represents the velocity tangential to the triangles edges. 

After substituting with the appropriate discrete operators, with $N=2$ for the 2D case, the discretization of Eq. \eqref{eq:vorticityNS}  takes the form       

\begin{multline}
\label{eq:discreteNS01}
[-\textrm{d}^T_0] \frac{U^{n+1} - U^n}{\Delta t} - \mu [-\textrm{d}^T_0] \ast_1 \textrm{d}_0 \ast^{-1}_0 \left[[-\textrm{d}^T_0] U + \textrm{d}_b V \right] \\ +  [-\textrm{d}^T_0] \ast_1 W_v \ast^{-1}_0 \left[[-\textrm{d}^T_0] U + \textrm{d}_b V\right] =0, 
\end{multline}
where $U$ is the vector containing the dual (normal) velocity 1-forms $u$ for all mesh dual edges, and the discrete wedge product of the tangential velocity 1-form $v$ with the 0-form $\ast \textrm{d} u$ is represented by the $W_v$ matrix, where the matrix $W_v$ contains the values of the tangential velocity 1-form $v$. The superscripts $n$ and $n+1$ in the time derivative term are due to time discretization and these subscripts are suppressed for the convective and viscous terms, deferring the time discretization of the viscous and convective terms until later in this subsection. The discrete operation $[-\textrm{d}^T_0 U]$ evaluates the circulation of the velocity forms $u$ along the dual 2-cells boundaries. Since a portion of these dual 2-cells boundaries may consists of primal edges, as discussed in Section \ref{subsec:DEC}, $[\textrm{d}_b V]$  then complements the velocity circulation, accounting for the part that depends on the velocity 1-forms $v$ on the primal boundary edges. The vector $V$ contains the tangential velocity 1-forms $v$ for all mesh primal edges, and the matrix $\textrm{d}_b$ is then defined as $\textrm{d}_b = 0.5 |\textrm{d}^T_0| diag(\textrm{d}^T_1 \mathbb{1}_2)$, where $|\textrm{d}^T_0|$ is the matrix $\textrm{d}^T_0$ with all entries made non-negative and $\mathbb{1}_2$ is a vector of ones with $N_2$ entries, and $diag(.)$ to be a diagonal matrix composed of the enclosed vector entries. The subscript $b$ emphasizes the fact that $\textrm{d}_b$ ``closes" the dual 2-cells for the boundary vertices by traversing along the primal boundary edges in the orientation direction of the dual 2-cells. Such a boundary contribution vanishes in the time derivative term  due to the time discretization. It also vanishes for the other $\textrm{d}^T_0$ operators in both the viscous and the convective terms, as shown in \ref{app:boundary-operator}.      

A definition for the discrete primal-primal wedge product was developed in \cite{hirani2003discrete} (pg. 74 Eq. (7.2.1)), according to which the wedge product between a discrete primal 1-form $\alpha$ and a discrete primal 0-form $\beta$ defined over a primal edge $[i,j]$ is 

\begin{equation}
\label{eq:discreteWedge2D}
\begin{aligned}
\langle \alpha \wedge \beta , [i,j] \rangle  &= \frac{1}{2} \left( \langle \alpha, [i,j] \rangle  \langle \beta, [j] \rangle - \langle \alpha, [j,i] \rangle  \langle \beta, [i] \rangle \right) \\ &= \frac{1}{2} \langle \alpha, [i,j] \rangle (  \langle \beta, [i] \rangle +  \langle \beta, [j] \rangle ),  
\end{aligned}
\end{equation}
where $[i,j]$ is the edge formed by nodes $[i]$ and $[j]$, $\langle \alpha, [i,j] \rangle$ is the discrete form $\alpha$ defined on the simplex $[i,j]$, with the property $\langle \alpha, [i,j] \rangle = - \langle \alpha, [j,i] \rangle$. Recalling that $\ast^{-1}_0 [-\textrm{d}^T_0] U$ is a vector with $N_0$ entries, $W_v$ is a sparse $N_1 \times N_0$ matrix defined as $W_v = 0.5 \ diag(V) |\textrm{d}_0|$. Accordingly, the action of this wedge product operation (in 2D) at each primal edge is to take the average of the vorticity $\ast^{-1}_0 [-\textrm{d}^T_0] U$ evaluated at the edge's nodes and multiply it by the edge's tangential velocity 1-form $v$. It is worth noting that the vector $U$ in Eq. \eqref{eq:discreteNS01} includes the normal velocity 1-forms $u$ for all mesh edges, including those that might be given as boundary conditions. 

In order to obtain a linear representation for the convective term, we consider the tangential velocity 1-forms $v$ to be given through an interpolation of previously-known normal velocity 1-forms $u$. A velocity vector field can be calculated inside each triangle through the interpolation of the velocity 1-forms $\ast_1^{-1} u$ defined on the triangle's faces. The interpolation is carried out here using Whitney maps \cite{HiNaCh2015}. Since the velocity 1-forms $\ast_1^{-1} u$ are closed forms; i.e. $\textrm{d}_1 \ast_1^{-1} u = 0$, the interpolation will result in a constant velocity vector field over each triangle (see \cite{arnold2010finite}, theorem 5.4). The constant vector field obtained by Whitney map interpolation is the one that would result from requiring a constant vector field with the given fluxes of an incompressible vector field. It is worth noting that other interpolation methods, like these in \cite{HallCavendish:1991dvmsvf}, can also be used to interpolate the velocity vector on triangles, which would result in velocity vectors not significantly different from these interpolated using Whitney maps. The tangential velocity 1-forms $v$ can be then calculated on each primal edge by averaging the velocity vector fields on the triangles sharing the edge. 

The second step in the exact fractional step method is to substitute for the velocity 1-forms $u$ (in Eq. \eqref{eq:discreteNS01}) by its definition as the curl of a stream function. The incompressibility condition (Eq. \eqref{eq:NS_DG-00-b}) in DEC notation is expressed as $\ast_2 \textrm{d}_1 \ast^{-1}_1 U$. This implies that the vector $U$ is in the null space of the matrix $[\ast_2 \textrm{d}_1 \ast^{-1}_1]$. Since $[\ast_2 \textrm{d}_1 \ast^{-1}_1] [\ast_1 \textrm{d}_0] = \ast_2 \textrm{d}_1 \textrm{d}_0 = 0$, the columns of the matrix $[\ast_1 \textrm{d}_0]$ then contain a basis of the null space of $[\ast_2 \textrm{d}_1 \ast^{-1}_1]$. Therefore, the vector $U$ can uniquely be expressed in terms of the basis $[\ast_1 \textrm{d}_0]$; i.e., $U = \ast_1 \textrm{d}_0 \Psi$. In vector calculus notation, this is equivalent to expressing a divergence-free velocity vector as the curl of a stream function. According to the diagram in Eq. \eqref{eq:cochncmplx2d}, the vector $\Psi$ includes the stream function $\psi$ defined as 0-forms on the primal mesh nodes (i.e. $\psi \in C^0(K)$). Substituting with this representation of $U$, Eq. \eqref{eq:discreteNS01} becomes    

\begin{multline}
\label{eq:discreteNS02}
\frac{1}{\Delta t} [-\textrm{d}^T_0] \ast_1 \textrm{d}_0 \Psi^{n+1} - \mu [-\textrm{d}^T_0] \ast_1 \textrm{d}_0 \ast^{-1}_0 [-\textrm{d}^T_0] \ast_1 \textrm{d}_0 \Psi \\ + [-\textrm{d}^T_0] \ast_1 W_v \ast^{-1}_0 [-\textrm{d}^T_0] \ast_1 \textrm{d}_0 \Psi =  F, 
\end{multline}
with the vector $F = \frac{1}{\Delta t} [-\textrm{d}^T_0] U^n + \mu [-\textrm{d}^T_0] \ast_1 \textrm{d}_0 \ast^{-1}_0  \textrm{d}_b V -  [-\textrm{d}^T_0] \ast_1 W_v \ast^{-1}_0  \textrm{d}_b V $. Eq. \eqref{eq:discreteNS02} describes the resultant linear system to be solved. The degrees of freedom in the above linear system are the stream function 0-forms (scalars) defined on the primal mesh nodes. Therefore, the resulting system is a sparse $N_0 \times N_0$ linear system.

For the current 2D case, it is worth noting the correspondence between the velocity 1-form $u$ and the mass flux across the primal edges. While the discrete velocity form $u$ is formally defined as the integration of the velocity field on the dual edges ($u=\int_{\star \sigma^1} \mathbf{u} . d \mathbf{l}$), it can be used to approximate the mass flux normal to the primal faces (edges) as $u_f = \ast_1^{-1} u$. The incompressibility condition then implies a zero summation of the mass fluxes across the faces (edges) of each triangular element; i.e., $\textrm{d}_1 u_f = \textrm{d}_1 [\ast^{-1}_1 u ] = 0$. On the other hand, if the mass flux across the primal edges is the known quantity, it can be used to approximate the velocity 1-form $u$ as $u = \ast_1 u_f$. Recalling that the actual degrees of freedom in the present discretization are the stream functions $\psi$ and the velocity 1-form $u$ is calculated as $u = \ast_1 \textrm{d}_0 \psi$, the mass flux across the primal edges is then $u_f = \textrm{d}_0 \psi$. This implies that it is the mass flux $u_f$ that is calculated first and is then used to approximate the velocity 1-form $u$ as $u = \ast_1 u_f = \ast_1 \textrm{d}_0 \psi$. Such a correspondence between the velocity 1-form $u$ and the mass flux $u_f$ provides some flexibility in the case that an initial analytical expression for the velocity vector field defined on a smooth surface is available. In order then to set the initial solution on the discrete triangulation objects, it is possible to calculate $u_f$ by integrating the mass flux normal to hypothetical curved primal edges and then approximate $u$ as $u = \ast_1 u_f$. Otherwise, the velocity 1-form $u$ can be initialized by integrating the velocity field on hypothetical curved dual edges. It is worth noting that the integration of the mass flux might be preferred in order to accurately guarantee zero net mass flux across the domain boundaries.    

We now elaborate on the time discretization wherein the linear system in Eq. \eqref{eq:discreteNS02} is solved in two steps as a predictor-corrector method. First, we advance the system explicitly by a half time step

\begin{multline}
\label{eq:discreteNS03}
\left[\frac{1}{0.5 \Delta t} [-\textrm{d}^T_0] \ast_1 \textrm{d}_0 \right] \Psi^{n+\frac{1}{2}} = F + \biggl[ \mu [-\textrm{d}^T_0] \ast_1 \textrm{d}_0 \ast^{-1}_0 [-\textrm{d}^T_0] \biggr. \\ \biggl. - [-\textrm{d}^T_0] \ast_1 W^n_v \ast^{-1}_0 [-\textrm{d}^T_0] \biggr] \ast_1 \textrm{d}_0 \Psi^n, 
\end{multline} 
where the matrix $W^n_v$ incorporates the tangential velocity forms $v$ at time $n$. After solving the linear system \eqref{eq:discreteNS03}, the normal velocity 1-forms at time $n+\frac{1}{2}$ are calculated as $U^{n+\frac{1}{2}} = \ast_1 \textrm{d}_0 \Psi^{n+\frac{1}{2}}$. These normal velocity 1-forms are then used to predict, at time step $n+\frac{1}{2}$, the velocity vector field at each triangle element through Whitney map interpolation \cite{HiNaCh2015}. The tangential velocity 1-form $v$ is then calculated at each primal edge as the average of the velocity vectors in the neighboring triangles, in the direction of the primal edge, multiplied by the primal edge length. This results in the tangential velocity 1-forms $v$ at time $n+\frac{1}{2}$, and then $W^{n+\frac{1}{2}}_v$. The latter matrix is then used to calculate the new velocity 1-forms $u$ at time $n+1$. The evaluation of the tangential velocity 1-forms $v$ at half time step ($n+\frac{1}{2}$) is expected to improve the kinetic energy conservation, as was shown before by Perot \cite{perot2000conservation} for the covolume method. The second linear system to solve is then

\begin{multline}
\label{eq:discreteNS04}
\biggl[ \frac{1}{\Delta t} [-\textrm{d}^T_0] \ast_1 \textrm{d}_0  - \mu [-\textrm{d}^T_0] \ast_1 \textrm{d}_0 \ast^{-1}_0 [-\textrm{d}^T_0] \ast_1 \textrm{d}_0  \biggr. \\ \biggl. + [-\textrm{d}^T_0] \ast_1 W^{n+\frac{1}{2}}_v \ast^{-1}_0 [-\textrm{d}^T_0] \ast_1 \textrm{d}_0 \biggr] \Psi^{n+1} = F, 
\end{multline}  
where the right hand side vector $F$ in both Eqs. \eqref{eq:discreteNS03} and \eqref{eq:discreteNS04} also contains the contribution from any stream function $\Psi$ boundary conditions.

The solution of the Navier-Stokes equations through the exact fractional step method is known to significantly reduce the size of the solved linear system \cite{HallCavendish:1991dvmsvf,chang2002analysis}. While the discretization of the momentum equation \eqref{eq:NS_DG04} would result in a number of degrees of freedom equals to the number of edges plus the number of triangles; i.e. ($N_1 + N_2$), the presented discretization has a number of degrees of freedom equals only to the number of primal nodes $N_0$. In addition to reducing the linear system size, the presented discretization always maintains the incompressibility error at the machine precision. Recalling the discrete incompressibility condition $\ast_2 \textrm{d}_1 \ast^{-1}_1 U = \ast_2 \textrm{d}_1 \ast^{-1}_1 \ast_1 \textrm{d}_0 \Psi = \ast_2 \textrm{d}_1 \textrm{d}_0 \Psi$, and since $\textrm{d}_1 \textrm{d}_0 = 0$, the resulting formulation guarantees the mass conservation up to the machine precision, regardless of the error incurred during the linear system solution. On the other hand, the solution of the Navier-Stokes equations in terms of a stream function might simplify or complicate the implementation of some boundary conditions. A discussion regarding the implementation of various boundary conditions, including the boundary conditions for interior walls (domain boundaries with zero in/out flow), through the stream function can be found in \cite{chang2002analysis}. For the case of interior boundaries with nonzero in/out flow, the boundary conditions can be implemented through Hodge decomposition manipulation for the stream functions. More details regarding the implementation of such boundary conditions might be addressed in future publications.     

The presented discretization of NS equations has similarities with previous discretizations.
The viscous term discretization is similar to that previously developed through the covolume method \cite{HallCavendish:1991dvmsvf,cavendish1992solution,choudhury1990discretization,nicolaides1993covolume,perot2000conservation} and DEC-based \cite{elcott2007stable,mullen2009energy} discretizations. However, the discretization of the convective term via the interior product definition in Eq. \eqref{eq:interior-product} and the wedge product definition in Eq. \eqref{eq:discreteWedge2D} makes the present discretization different from previous DEC-based discretizations that adopted Lie derivative advection techniques \cite{elcott2007stable} or a finite-volume-based approach \cite{mullen2009energy} in discretizing the advection term. It also distinguishes the present discretization from most of the covolume method discretizations, with some similarity with the covolume discretization developed by Perot \cite{perot2000conservation} only in the special case of structured-triangular meshes; i.e. when the center points of the primal edges and their dual edges coincide. Nevertheless, unlike all covolume discretizations, the current discretization is capable of simulating flows over both flat and curved surfaces. In addition, the current manipulation of the convective term, through the discrete wedge product operator, gives insight into other important research themes. For example, it paves the way to exploring the discretization of Navier-Stokes equations through the finite element exterior calculus method \cite{arnold2010finite,arnold2006finite}. Furthermore, it gives insight into the discretization of other convective terms in different physics problems; e.g. the magnetohydrodynamics governing equations.

\subsection{Three dimensional discretization}
\label{subsec:discretization:3D}

The three-dimensional DEC discretization of NS equations is briefly presented in this subsection. In the 3D situation, the primal mesh consists of 3-simplices (tetrahedra), 2-simplices (triangles), 1-simplices (edges) and 0-simplices (nodes). The duals to these primal mesh entities consist of 0-cells (dual nodes), 1-cells (dual edges), 2-cells (polygons) and 3-cells (polyhedra), respectively. The space of the discrete $k$-forms defined on primal/dual mesh complexes is defined according to the diagram in Eq. \eqref{eq:cochncmplx3d}.  

Following the same methodology in Section \ref{subsec:discretization:2D}, we start the discretization of Eq. \eqref{eq:vorticityNS} by choosing to define the 2-forms $\textrm{d} \mathbf{u}^\flat$, in the time derivative term, on the dual 2-cells (polygons). It then follows that the velocity 1-forms $\mathbf{u}^\flat$ in this term are defined on the dual 1-cells (i.e. dual edges). The velocity 1-forms defined on the dual edges are denoted by $u$. It also follows, based on the diagram in Eq. \eqref{eq:cochncmplx3d}, that the velocity 1-forms in the viscous term are also defined on the dual edges. In regards to the convective term, since $\textrm{d} \ast(\mathbf{u}^\flat \wedge \ast \textrm{d} \mathbf{u}^\flat)$  is defined on the dual 2-cells, then $\ast(\mathbf{u}^\flat \wedge \ast \textrm{d} \mathbf{u}^\flat)$ has to be defined on the dual 1-cells (dual edges), and therefore $(\mathbf{u}^\flat \wedge \ast \textrm{d} \mathbf{u}^\flat)$ is defined on the primal $2$-simplices (triangles). Since $\mathbf{u}^\flat$ is a 1-form, then $\ast \textrm{d} \mathbf{u}^\flat$ is also 1-form (in the 3D case), and therefore the wedge product is carried out between two 1-forms, where the outcome of this wedge product is defined on the primal triangles. This implies that both the wedge product portions are defined on the primal edges. Accordingly, the 1-form $\ast \textrm{d} \mathbf{u}^\flat$ is defined on the primal edges, which makes $\mathbf{u}^\flat$ in this portion to be defined on the dual edges. Finally, the first velocity 1-form in the wedge product term is defined on the primal edges. We denote the velocity 1-forms defined on the primal edges by $v$. 

Substituting with the appropriate discrete operators, with $N=3$ for the 3D case, the discretization of Eq. \eqref{eq:vorticityNS} then takes the form       

\begin{equation}
\label{eq:discreteNS_3D_1}
 \textrm{d}^T_1 \frac{U^{n+1} - U^n}{\Delta t} + \mu \textrm{d}^T_1 \ast_2 \textrm{d}_1 \ast^{-1}_1 \textrm{d}^T_1 U -  \textrm{d}^T_1 \ast_2 W_v \ast^{-1}_1 \textrm{d}^T_1 U =0, 
\end{equation}
where $U$ is a vector containing the dual (normal) velocity 1-forms $u$ for all mesh dual edges, and $W_v$ is a sparse matrix representing the action of the discrete wedge product operator. It is worth noting that similar to Eq \eqref{eq:discreteNS01}, the operator $\textrm{d}^T_1$ needs to be complemented by another operator that closes some of the dual 2-cells on the domain boundary. Such a complementary matrix is omitted in Eq. \eqref{eq:discreteNS_3D_1} for simplicity, but its existence should always be considered during numerical implementations.  

According to the definition in \cite{hirani2003discrete} (pg. 74 Eq. (7.2.1)), the wedge product of a discrete primal 1-form $\alpha$ and a discrete primal 1-form $\beta$ defined over a primal 2-simplex (triangle) $[i,j,k]$ is 

\begin{equation}
\label{eq:discreteWedge3D}
\begin{aligned}
\langle \alpha \wedge \beta , [i,j,k] \rangle  = \frac{1}{6} (  &\langle \alpha, [i,j] \rangle  \langle \beta, [j,k] \rangle - \langle \alpha, [i,k] \rangle  \langle \beta, [k,j] \rangle \\ 
- & \langle \alpha, [j,i] \rangle  \langle \beta, [i,k] \rangle  + \langle \alpha, [j,k] \rangle  \langle \beta, [k,i] \rangle \\
+ & \langle \alpha, [k,i] \rangle  \langle \beta, [i,j] \rangle -\langle \alpha, [k,j] \rangle  \langle \beta, [j,i] \rangle ),  
\end{aligned}
\end{equation}
with the property $\langle \alpha, [i,j] \rangle = - \langle \alpha, [j,i] \rangle$. Recalling that $\ast^{-1}_1 \textrm{d}^T_1 U$ is a vector of $N_1$ rows, $W_v$ is then a sparse $N_2 \times N_1$ matrix. For each row in $W_v$ corresponding to a primal triangle, the only non-zero entries in this row are for the primal edges belonging to this triangle. For a primal triangle $[i,j,k]$, the matrix entry corresponding to the edge $[i,j]$ is $\frac{1}{6} ( \langle v, [k,i] \rangle - \langle v, [j,k] \rangle )$.

Following the procedure in section \ref{subsec:discretization:2D}, the velocity 1-forms are substituted by their definition in terms of a stream function; i.e. $U = \ast_2 \textrm{d}_1 \Psi$. According to the diagram in Eq. \eqref{eq:cochncmplx3d}, the vector $\Psi$ includes the stream functions $\psi$ defined as 1-forms on the primal mesh edges ($\psi \in C^1(K)$). Substituting in Eq. \eqref{eq:discreteNS_3D_1}, the resulting linear system is    

\begin{equation}
\label{eq:discreteNS_3D_2}
 \textrm{d}^T_1 \frac{1}{\Delta t} \ast_2 \textrm{d}_1 \Psi^{n+1} + \mu \textrm{d}^T_1 \ast_2 \textrm{d}_1 \ast^{-1}_1 \textrm{d}^T_1 \ast_2 \textrm{d}_1 \Psi -  \textrm{d}^T_1 \ast_2 W_v \ast^{-1}_1 \textrm{d}^T_1 \ast_2 \textrm{d}_1 \Psi = \textrm{d}^T_1 \frac{1}{\Delta t} U^n.  
\end{equation}

The degrees of freedom in the above linear system are the stream function 1-forms (scalars) defined on the primal edges. Therefore, the resulting system is a sparse $N_1 \times N_1$ linear system. The time discretization of Eq. \eqref{eq:discreteNS_3D_2} can then be implemented similar to the 2D case.

The presented 3D discretization of the viscous term in Eq. \eqref{eq:discreteNS_3D_2} is similar to previous DEC-based \cite{elcott2007stable,mullen2009energy} and covolume method \cite{cavendish1994complementary,zhang2002accuracy} discretizations. However, using the interior product definition in Eq. \eqref{eq:interior-product} and the discrete wedge product definition in Eq. \eqref{eq:discreteWedge3D} makes the present discretization different from all previously developed 3D DEC-based/covolume discretizations.           
 
\section{Results and Discussion}
\label{sec:results}

In order to benchmark the performance of the presented discretization, several 2D simulation experiments are performed for flows over both flat and curved surfaces. During all simulations, Eqs. \eqref{eq:discreteNS03} and \eqref{eq:discreteNS04} are solved consecutively at each time step, where direct LU decomposition solver is used to solve the linear systems. As pointed out earlier, the mass conservation is guaranteed by the discretization construction. Vorticity is also locally conserved due to the discretization construction, as shown earlier by Perot \cite{perot2000conservation}. Global vorticity conservation up to the machine precision was observed during all conducted simulations, where in the presence of solid walls the vorticity flux comes only from the no slip boundaries as it should be for incompressible flows. Therefore, the results presented below mainly quantify the numerical convergence rate of the discretization and the conservation of the kinetic energy.
        

\subsection{Driven cavity}
\label{subsec:Driven-cavity}
Driven cavity simulations are carried out on a unit square domain at Reynolds number (Re) of 1000. Solid wall boundary conditions are imposed on the left, right and bottom boundaries. The top boundary has zero flux (i.e. $u$), and a unit tangential velocity (i.e. $v$) boundary conditions. Therefore, the stream functions on all boundary nodes are set to an arbitrary constant value. The fluid dynamic viscosity ($\mu = 1/Re$ in our normalized units) is set to $0.001$, and the time step is $\Delta t = 0.1$. The simulations are carried out on a Delaunay mesh and a structured-triangular mesh (consisting of isosceles right triangles) with 32482 and 32258 elements, respectively, which has almost the same resolution as a $128 \times 128$ Cartesian mesh.

\begin{figure}
\begin{subfigure}[b]{0.5\textwidth}
\centering
\includegraphics[width=\textwidth]{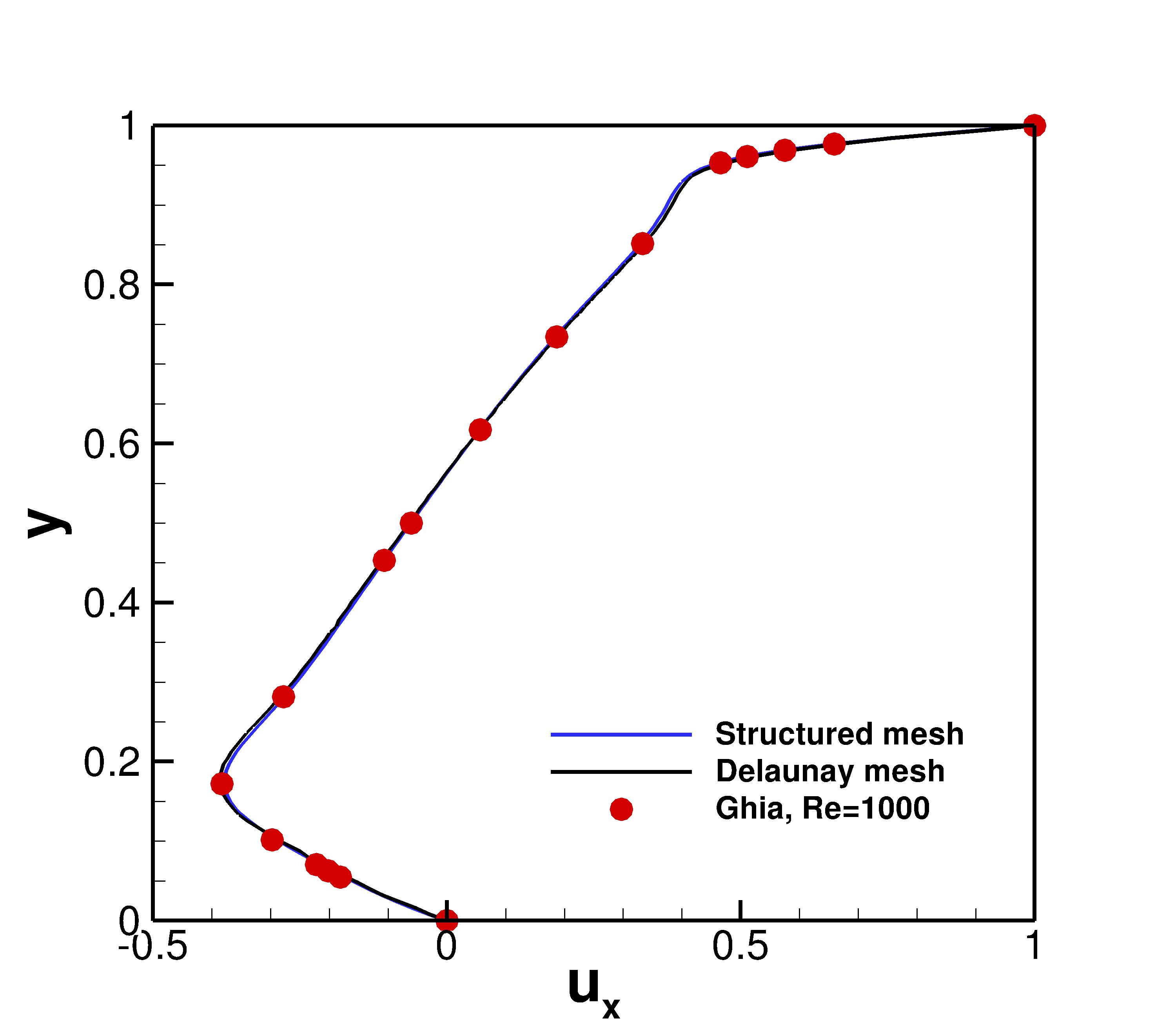}
\caption{}
\end{subfigure} %
\begin{subfigure}[b]{0.5\textwidth}
\centering
\includegraphics[width=\textwidth]{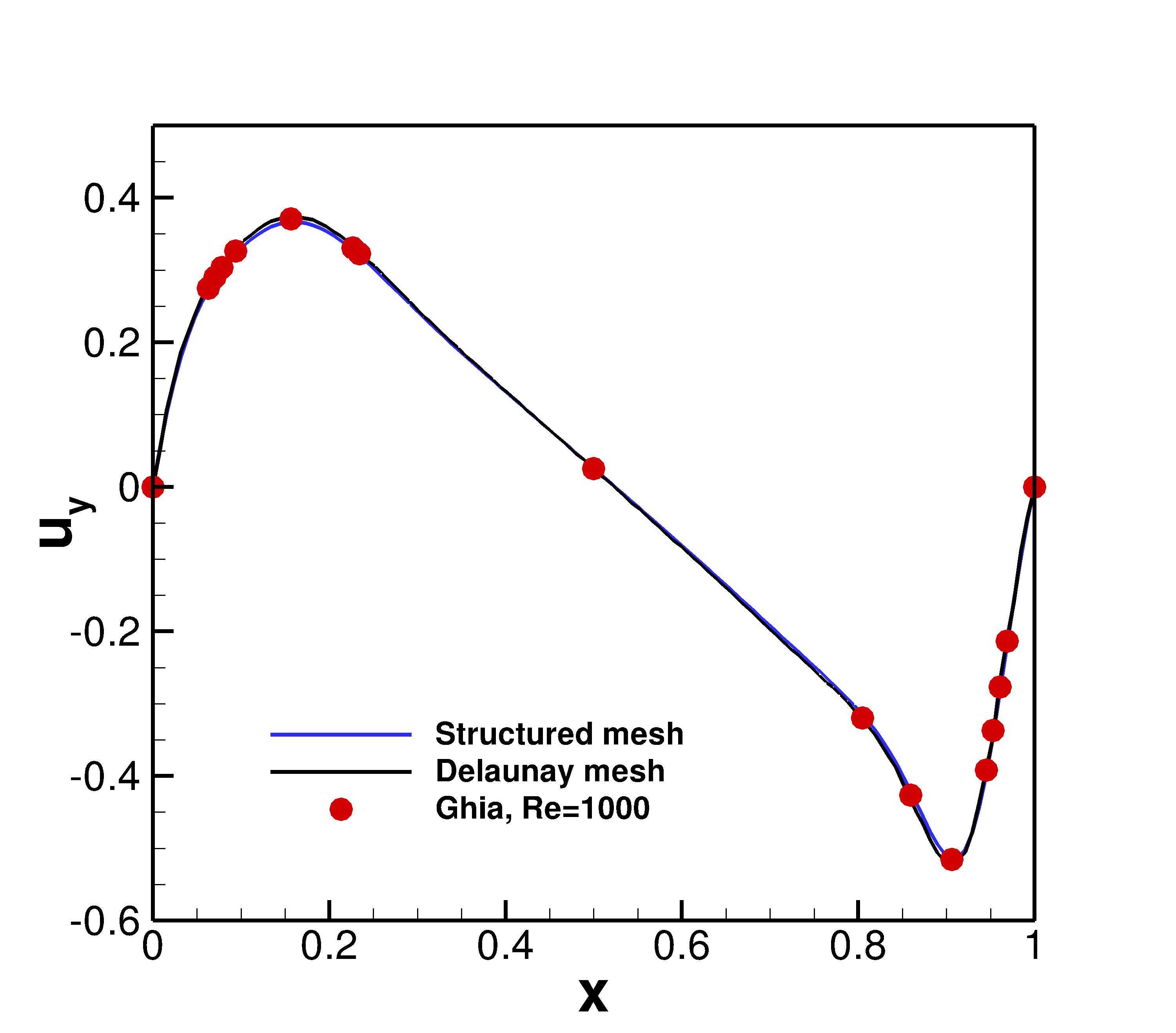} 
\caption{}
\end{subfigure} %
\caption{Cross-section of the velocity profile at the two domain center lines for driven cavity test case at Reynolds number = 1000.}
\label{Fig:Driven-cavity}
\end{figure}

Fig. \ref{Fig:Driven-cavity} shows cross-sections at the domain center lines for the steady state velocity profile at simulation time $T=100$. The results are compared with well-established simulations by Ghia \textit{et al.} \cite{ghia1982high} for Re = 1000 using a $128 \times 128$ Cartesian mesh. The comparison shows an agreement with Ghia's results for both mesh types, which reflects the numerical solution fidelity.              

\subsection{Taylor-Green vortices}
\label{subsec:Taylor-Green-vortices}

The simulation of Taylor-Green vortices is carried out on a square domain of dimension $[-\pi, \pi]$ in both $x$ and $y$ directions. The decay of Taylor-Green vortices with time has an analytical solution that for the 2D case is expressed as \cite{green1937mechanism,connors2010convergence}
\begin{equation}
\begin{aligned}
& u_x = -\cos(x) \sin(y) e^{-2 \nu t}   \\
& u_y =  \sin(x) \cos(y) e^{-2 \nu t}
\end{aligned} 
\label{Eq:Taylor-Green-vortices}
\end{equation}   
with $\nu$ to be the kinematic viscosity. The simulation is conducted using a Delaunay mesh consisting of $50852$ elements with periodic boundary conditions applied on all domain boundaries. This requires only to fix the stream function at one primal node to an arbitrary value in order to get a unique solution. The simulation is carried out using a time step $\Delta t = 0.1$ and kinematic viscosity $\nu = 0.01$.

\begin{figure}
\begin{subfigure}[b]{0.5\textwidth}
\centering
\includegraphics[width=\textwidth]{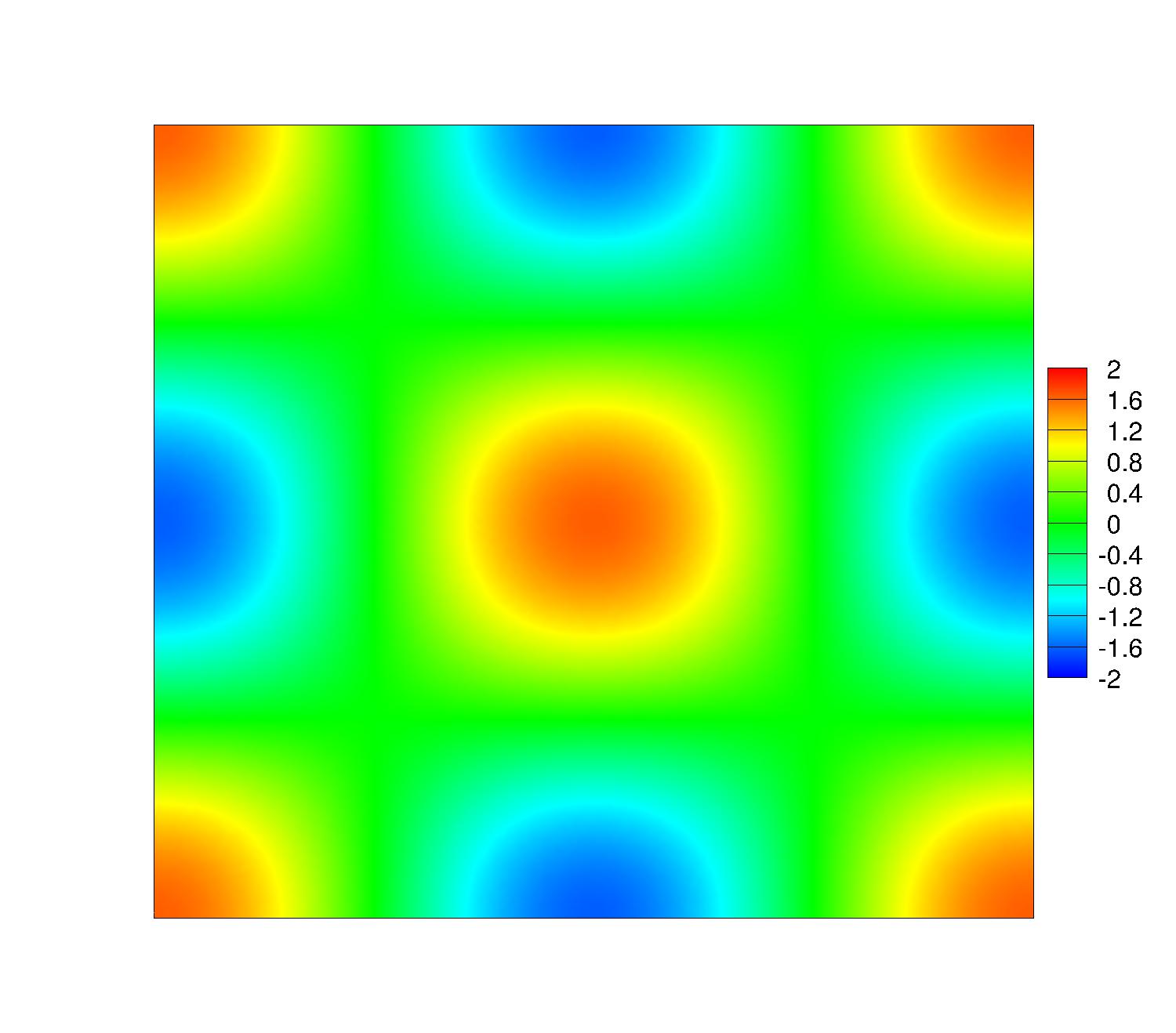}
\caption{}
\label{Fig:Taylor-Green-vortices-a}
\end{subfigure} %
\begin{subfigure}[b]{0.5\textwidth}
\centering
\includegraphics[width=\textwidth]{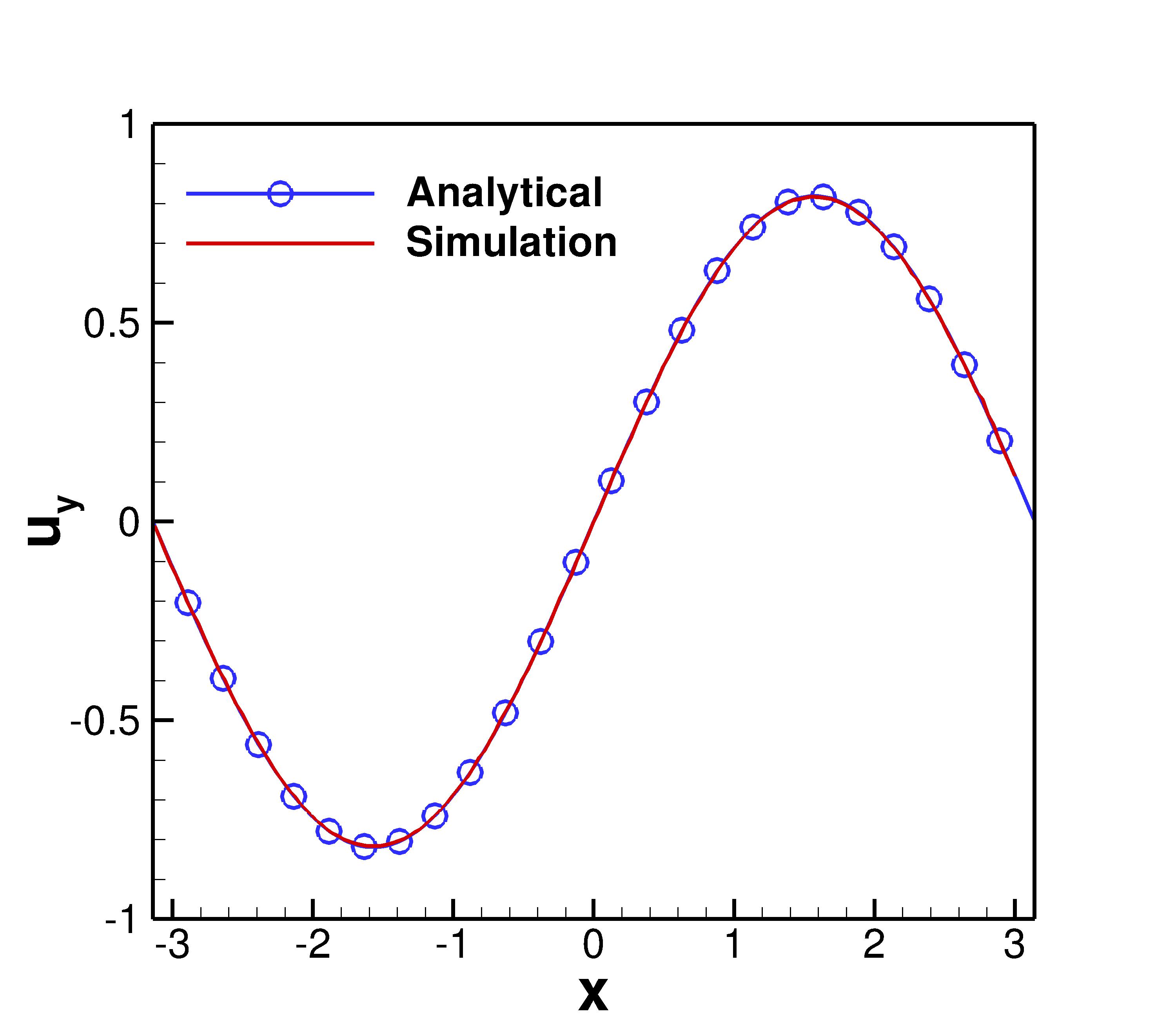} 
\caption{}
\label{Fig:Taylor-Green-vortices-b}
\end{subfigure} %
\caption{ (\subref{Fig:Taylor-Green-vortices-a}) The vorticity contour plot for Taylor-Green vortices at time $T=10$. (\subref{Fig:Taylor-Green-vortices-b}) Cross-section of the velocity y-component profile at the horizontal center line for Taylor-Green vortices at time $T=10$.}
\label{Fig:Taylor-Green-vortices}
\end{figure}

Fig. \ref{Fig:Taylor-Green-vortices-a} shows the vorticity contour plot for Taylor-Green vortices at simulation time $T=10$. A cross section of the velocity y-component $u_y$ along the horizontal domain center line is shown in Fig. \ref{Fig:Taylor-Green-vortices-b}. The simulation velocity profile is in good agreement with the analytical solution, as shown in Fig. \ref{Fig:Taylor-Green-vortices-b}. This represents a qualitative indication of the reliability of the current numerical implementation to reproduce the evolution of such unsteady flow with time.    

\subsection{Poiseuille flow}
\label{subsec:Poiseuille-flow}
Poiseuille flow simulations are carried out to investigate the numerical convergence rate of the developed discretization. The simulations are conducted on a unit square domain. Solid wall boundary conditions are imposed on the top and bottom boundaries, while parabolic in/out flow conditions are imposed on the left/right boundaries. Therefore, by fixing the stream function at one boundary node, the stream functions at the rest of the boundary nodes can be directly calculated based on the in/out flux (i.e. $u$) boundary conditions. The simulation is carried out for structured-triangular, Delaunay and well-centered meshes of different resolutions. The well-centered mesh is a Delaunay mesh that is optimized to make the circumcenter of each triangle to reside inside the triangle itself \cite{VaHiGuRa2010}.  

\begin{figure}
\centering
\includegraphics[width=0.6\textwidth]{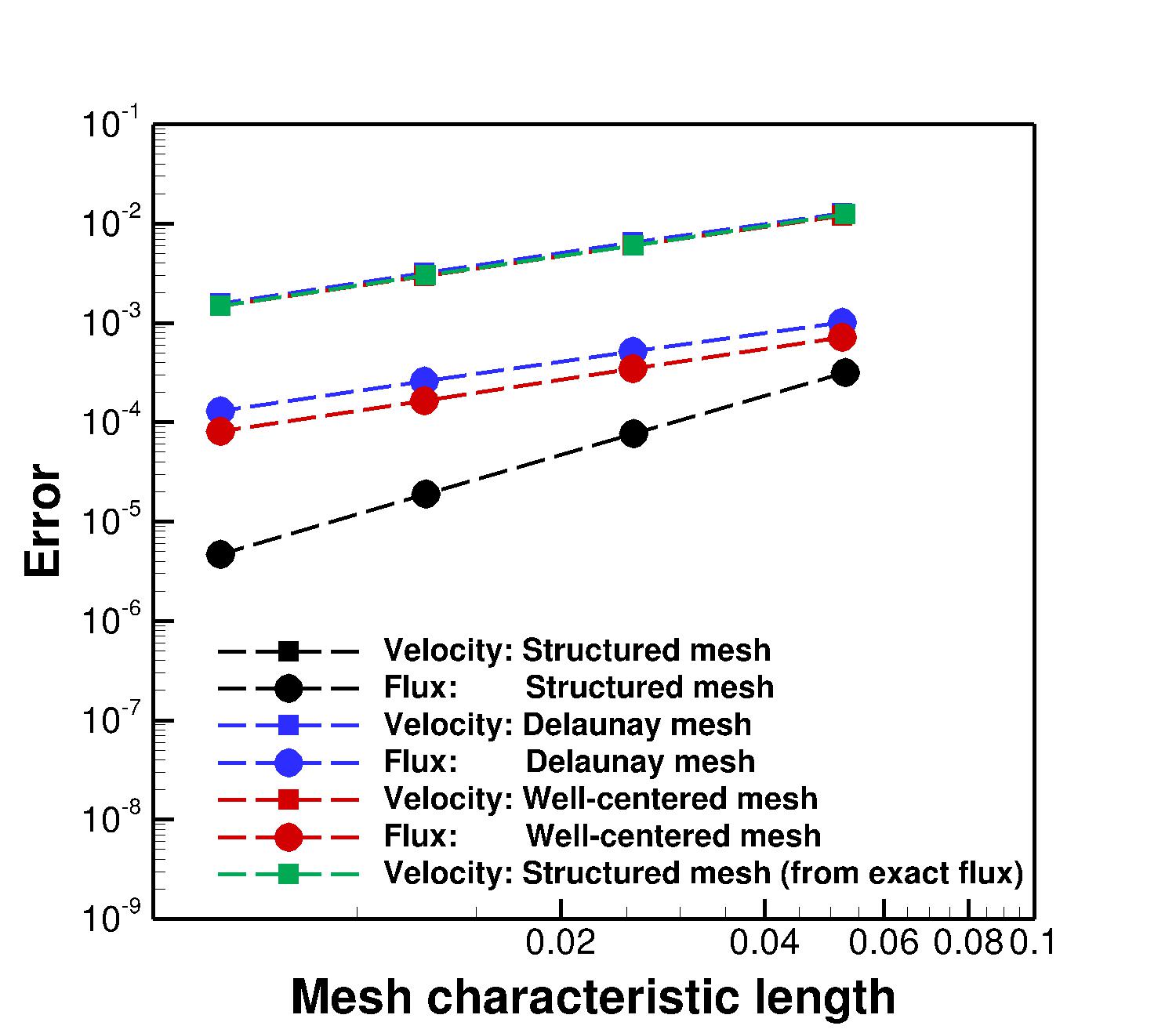}
\caption{The numerical convergence of the velocity 1-form (flux) and the interpolated velocity vector for the Poiseuille flow test case. The dashed lines represent the 1-st and 2-nd order slopes.}
\label{Fig:PoiseulleConvergence}
\end{figure}

The exact solution of the velocity vector field is given by $\mathbf{u} = [y(1-y),0]$. The $L^2$-norm of the velocity 1-form ($u$) error (see Hall \textit{et al.} \cite{HallCavendish:1991dvmsvf}) is calculated as $\| u^{exact} - u\| = \left[ \sum_{\sigma^1} \left(\frac{u^{exact} - u}{|\sigma^1|}  \right)^2 |\sigma^1| \ |\star \sigma^1| \right]^{1/2}$, and its convergence with the mesh elements size is shown in Fig. \ref{Fig:PoiseulleConvergence}. It is observed that the velocity 1-form (flux) error converges with a second order rate for the structured-triangular mesh case, and with a first order rate otherwise. This is in agreement with previous theoretical analysis by Nicolaides \cite{nicolaides:1992direct} for the covolume method. Such analyses showed that a necessary condition to obtain a second order convergence rate is to have the midpoint of each primal edge to coincide with the midpoint of its dual edge, which is satisfied only for a structured-triangular mesh or a mesh consisting of equilateral triangles. The observed convergence rates are therefore in agreement with theory. Regarding the unstructured meshes, the well-centered mesh error is slightly smaller than the Delaunay mesh error, where both the well-centered and Delaunay mesh implementations converge in a first order fashion.       

The convergence of the interpolated velocity vector field is also investigated. The velocity vector field is calculated inside each triangle through the interpolation of the velocity 1-forms $\ast_1^{-1} u$ defined on the triangle's faces using Whitney maps \cite{HiNaCh2015}. As pointed out earlier in section \ref{subsec:discretization:2D}, such interpolation results in a constant velocity vector field over each triangle, implying a first order interpolation scheme. The velocity field is interpolated over all triangles and the $L^2$-norm of the velocity vector error is calculated. The convergence of the velocity vector error with the mesh size is shown in Fig. \ref{Fig:PoiseulleConvergence}. A first order convergence rate is observed for all considered mesh types. Although the structured-triangular mesh exhibited a second order convergence for the flux 1-forms, the interpolated velocity vector converges with a first order rate. This can be attributed to the first order velocity interpolation scheme, which seems to dominate the velocity vector error. This is confirmed by calculating the velocity vector through the interpolation of the exact flux 1-forms (calculated by integrating the velocity analytical solution over the dual edges), which also converges with a first order rate, as shown in Fig. \ref{Fig:PoiseulleConvergence}.    


\subsection{Double periodic shear layer}
\label{subsec:shear-layer}
The simulation of a double periodic shear layer is carried out for an inviscid flow ($\mu = 0$) over a square domain of unit edge length. The initial flow represents a shear layer of finite thickness with a small magnitude of vertical velocity perturbation. The initial velocity vector field is expressed as \cite{bell1989second} 

\begin{equation}
\label{eq:double-shear-layer}
\begin{aligned}
u_x &=\begin{cases}
    \tanh ((y-0.25)/\rho), & \text{for $y \leq 0.5$},\\
    \tanh ((0.75-y)/\rho), & \text{for $y > 0.5$},
  \end{cases} \\
u_y &= \delta \sin(2\pi x),   
\end{aligned}
\end{equation}
with $\rho = 1/30$ and $\delta = 0.05$. The initial velocity 1-forms $u$ are approximated by integrating mass flux normal to the primal edges and then multiplying this flux by the discrete Hodge star operator $\ast_1$. Periodic boundary conditions are imposed on all domain boundaries. Therefore, it is only required to fix the stream function at one primal node to an arbitrary value in order to get a unique solution. 

\begin{figure}
\begin{subfigure}[b]{0.5\textwidth}
\centering
\includegraphics[width=\textwidth]{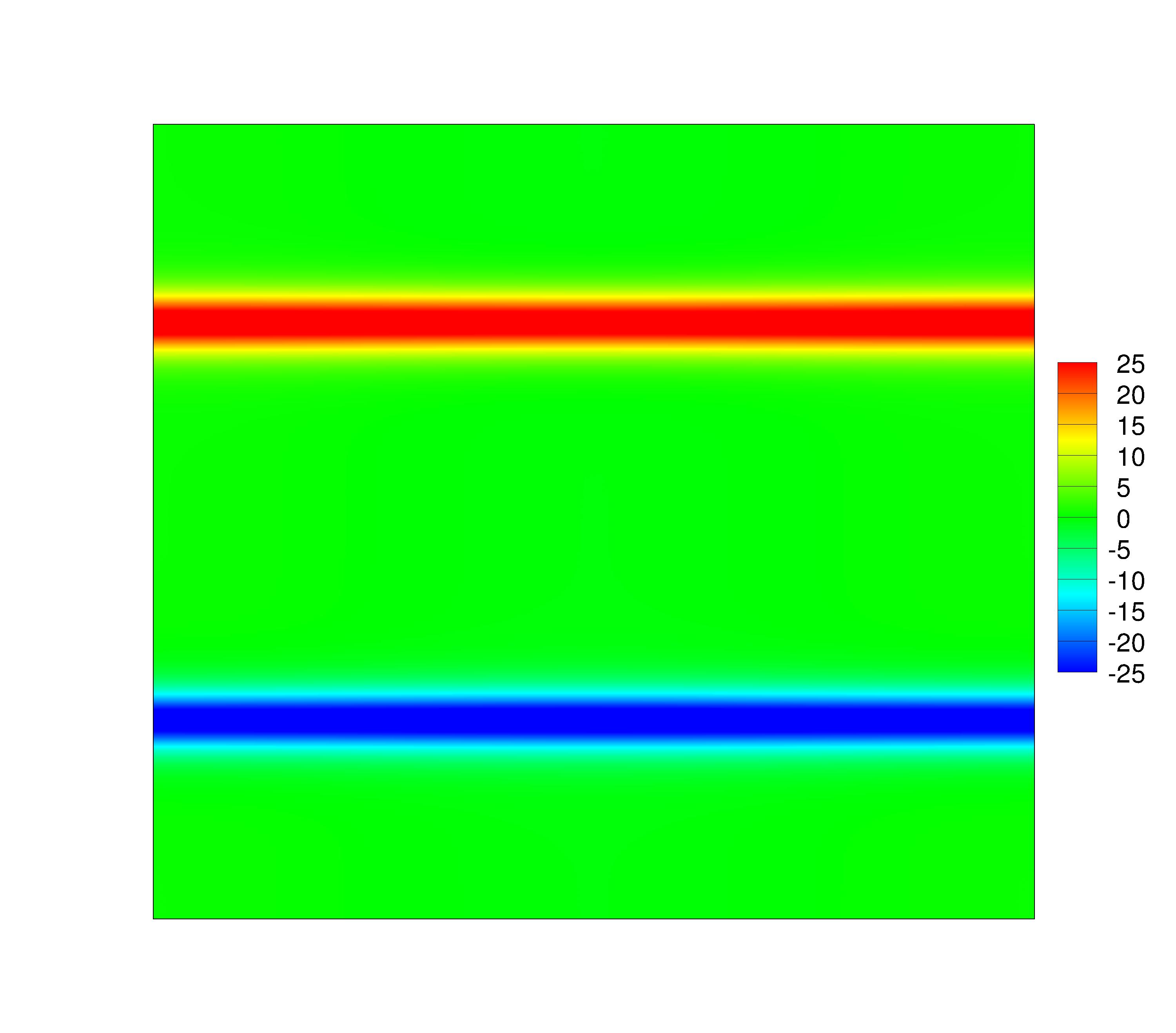}
\caption{}
\label{Fig:Double-shear-layer-a}
\end{subfigure} %
\begin{subfigure}[b]{0.5\textwidth}
\centering
\includegraphics[width=\textwidth]{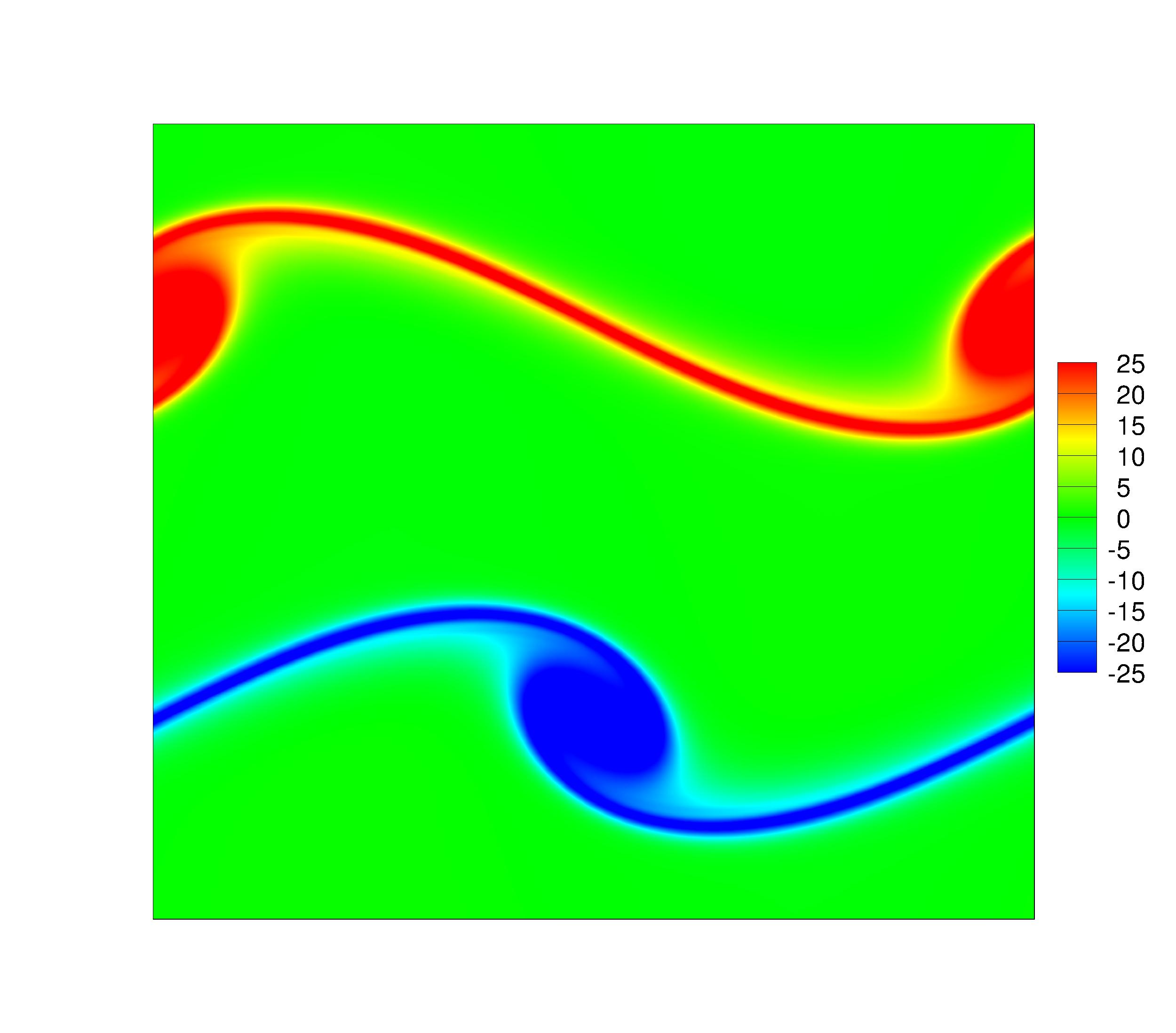} 
\caption{}
\label{Fig:Double-shear-layer-b}
\end{subfigure} %
\begin{subfigure}[b]{0.5\textwidth}
\centering
\includegraphics[width=\textwidth]{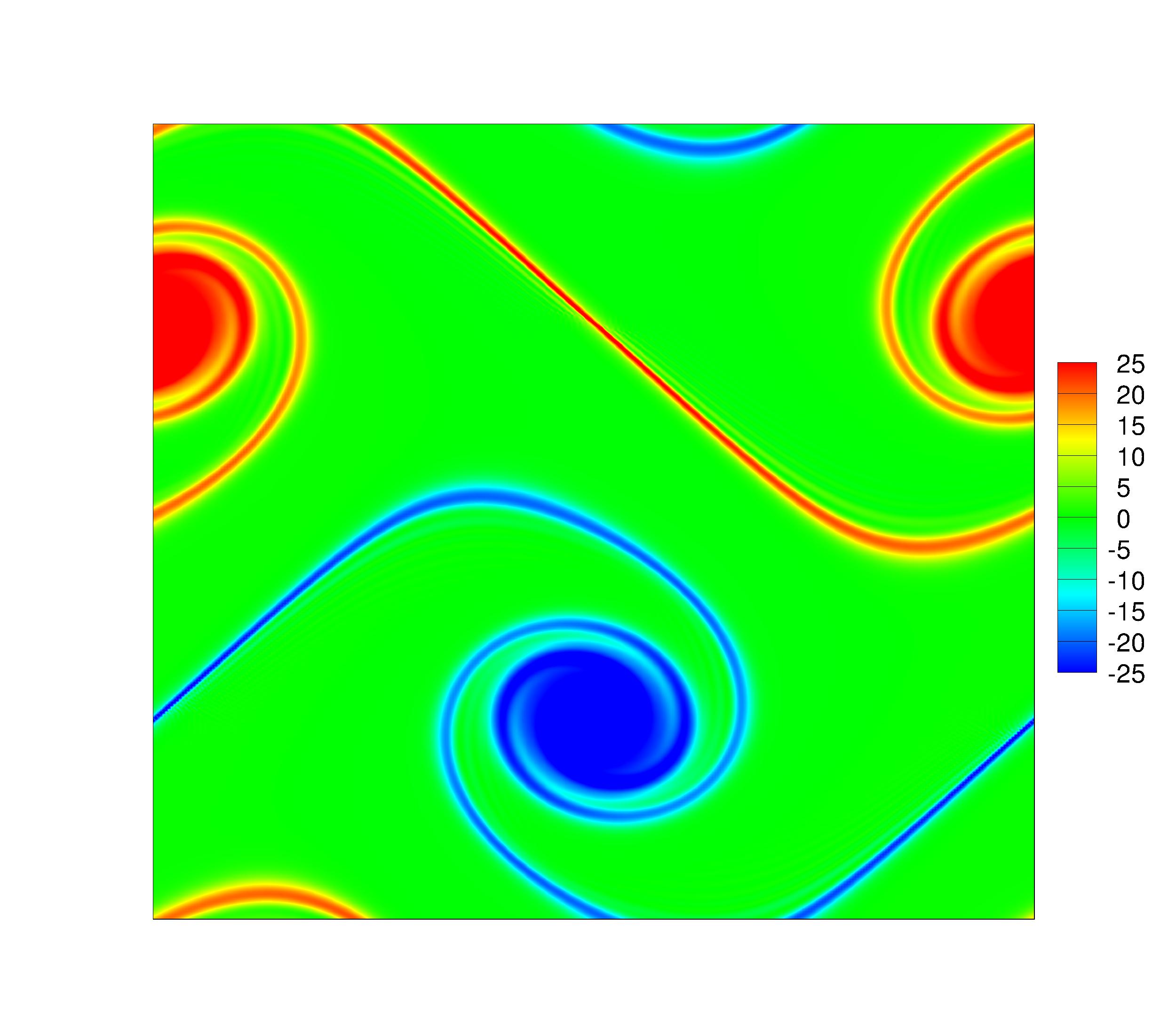}
\caption{}
\label{Fig:Double-shear-layer-c}
\end{subfigure} %
\begin{subfigure}[b]{0.5\textwidth}
\centering
\includegraphics[width=\textwidth]{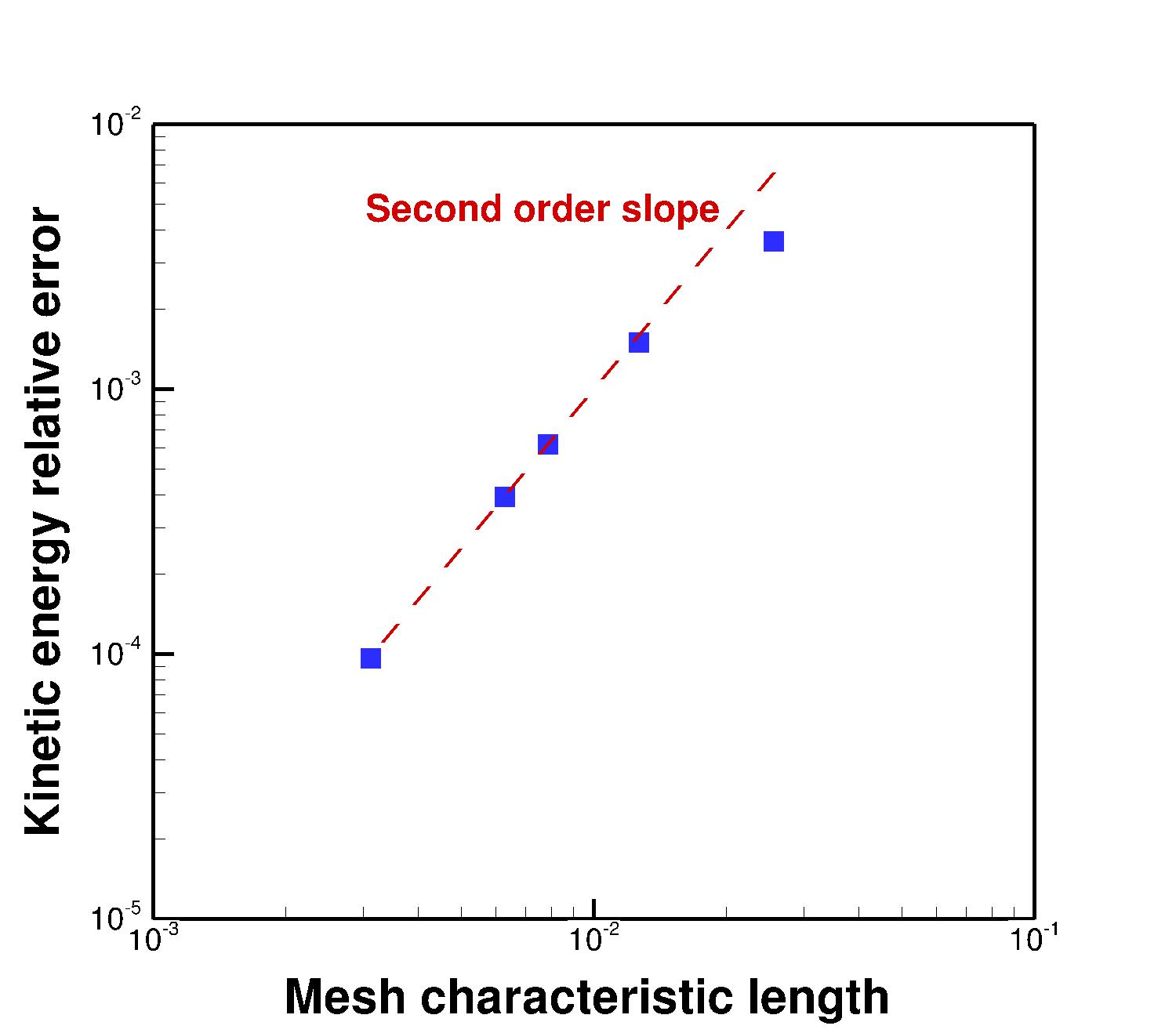} 
\caption{}
\label{Fig:Double-shear-layer-d}
\end{subfigure} %
\caption{The vorticity contour plot for double periodic shear layer with a mesh of 204800 elements at time: (\subref{Fig:Double-shear-layer-a}) T=0.0, (\subref{Fig:Double-shear-layer-b}) T=0.8 and (\subref{Fig:Double-shear-layer-c}) T=1.2. (\subref{Fig:Double-shear-layer-d}) The convergence of the relative kinetic energy error ($\frac{KE(0) - KE(T)}{KE(0)}$) with the characteristic mesh length at simulation time $T=2.0$. The dashed red line represents the second order slope.}
\label{Fig:Double-shear-layer}
\end{figure}

Five simulations are conducted using a time step of $\Delta t = 0.001$ on structured-triangular meshes with number of elements equal to 3042, 12482, 32258, 50562 and 204800. Fig. \ref{Fig:Double-shear-layer} shows the evolution of the vorticity contour plot with time, using the finest mesh. At time $T=0.8$, two vortices appear to be well resolved. The shear layer connecting the coherent vortices becomes thinner with time and within a finite time interval reach the resolution of the mesh after which dispersion error is manifested as mesh level oscillations. The vorticity contour plot in Fig. \ref{Fig:Double-shear-layer} exhibit similarities with previous simulations by Bell \textit{et al.} \cite{bell1989second}.

The convergence of the kinetic energy error with the mesh size is investigated.  The kinetic energy is calculated as $ \int_{\Omega} \mathbf{u} . \mathbf{u} \ d \Omega $, where the integration is carried out over the entire simulation domain. The velocity vector is calculated in each triangular element via Whitney map interpolation, as described before. The kinetic energy relative error ($\frac{KE(0) - KE(T)}{KE(0)}$) is then calculated at time $T=2.0$ and plotted versus the mesh characteristic length in Fig. \ref{Fig:Double-shear-layer-d}. Except for the coarsest mesh case, the kinetic energy relative error converges in a second order fashion with the mesh size, which is expected from a scheme that is second order for structured-triangular meshes. Overall, the kinetic energy relative error is modest, with a $0.3 \%$ error for the coarsest mesh (equivalent to a $40 \times 40$ Cartesian mesh) and only $0.01 \%$ error for the finest mesh (equivalent to a $320 \times 320$ Cartesian mesh). For the mesh with 50562 triangular elements (equivalent to a $128 \times 128$ Cartesian mesh), the kinetic energy relative error is $0.039\%$, almost one order of magnitude lower than a second order collocated mesh scheme using almost the same mesh size \cite{bell1989second}.


\subsection{Taylor vortices on flat surfaces}
\label{subsec:flat-Taylor-vortices}

Two Taylor vortices are simulated for an inviscid flow ($\mu = 0$) over a flat square domain of dimension $[-\pi, \pi]$ in both directions. The vorticity distribution for each vortex is expressed as \cite{mckenzie2007hola}

\begin{equation}
\label{eq:taylor-vorticity-distribution}
\omega(x,y) = \frac{G}{a} \left( 2-\frac{r^2}{a^2} \right) \exp \left( 0.5 \left( 1-\frac{r^2}{a^2} \right)  \right),
\end{equation}
with $G=1.0$, $a=0.3$ and $r$ is the distance between any field point and the vortex center. The vorticity distribution in Eq. \eqref{eq:taylor-vorticity-distribution} ensures that the net circulation of each Taylor vortex is zero.

The domain is initialized with a vorticity distribution for two vortices separated by a distance of $0.8$. Such a separation distance is just above the critical bifurcation distance, below which the two vortices would merge \cite{mckenzie2007hola,pavlov2011structure}. The vorticity values are assigned to the primal nodes according to Eq. \eqref{eq:taylor-vorticity-distribution}. The velocity 1-forms $u$ are determined by solving the Poisson equation

\begin{equation}
\label{eq:taylor-vorticity-Poisson}
\ast^{-1}_0 \textrm{d}^T_0 \ast_1 \textrm{d}_0 \Psi = X,
\end{equation}
where $X$ is the vector containing the known vorticity values, and $\Psi$ is the vector containing the unknown stream functions on the primal nodes. No-flux Dirichlet boundary conditions are imposed on the domain boundaries during the Poisson equation solution. 

The Poisson equation is solved only once initially and the velocity 1-forms are then calculated as $U=\ast_1 \textrm{d}_0 \Psi$. Such velocity 1-forms are used as the initial state for the simulations. When simulating the evolution of the two Taylor vortices, periodic boundary conditions are imposed on all domain boundaries. Therefore, it is only required to fix the stream function at one primal node to an arbitrary value in order to get a unique solution. 

\begin{figure}
\centering
\begin{subfigure}[b]{0.4\textwidth}
\includegraphics[width=\textwidth]{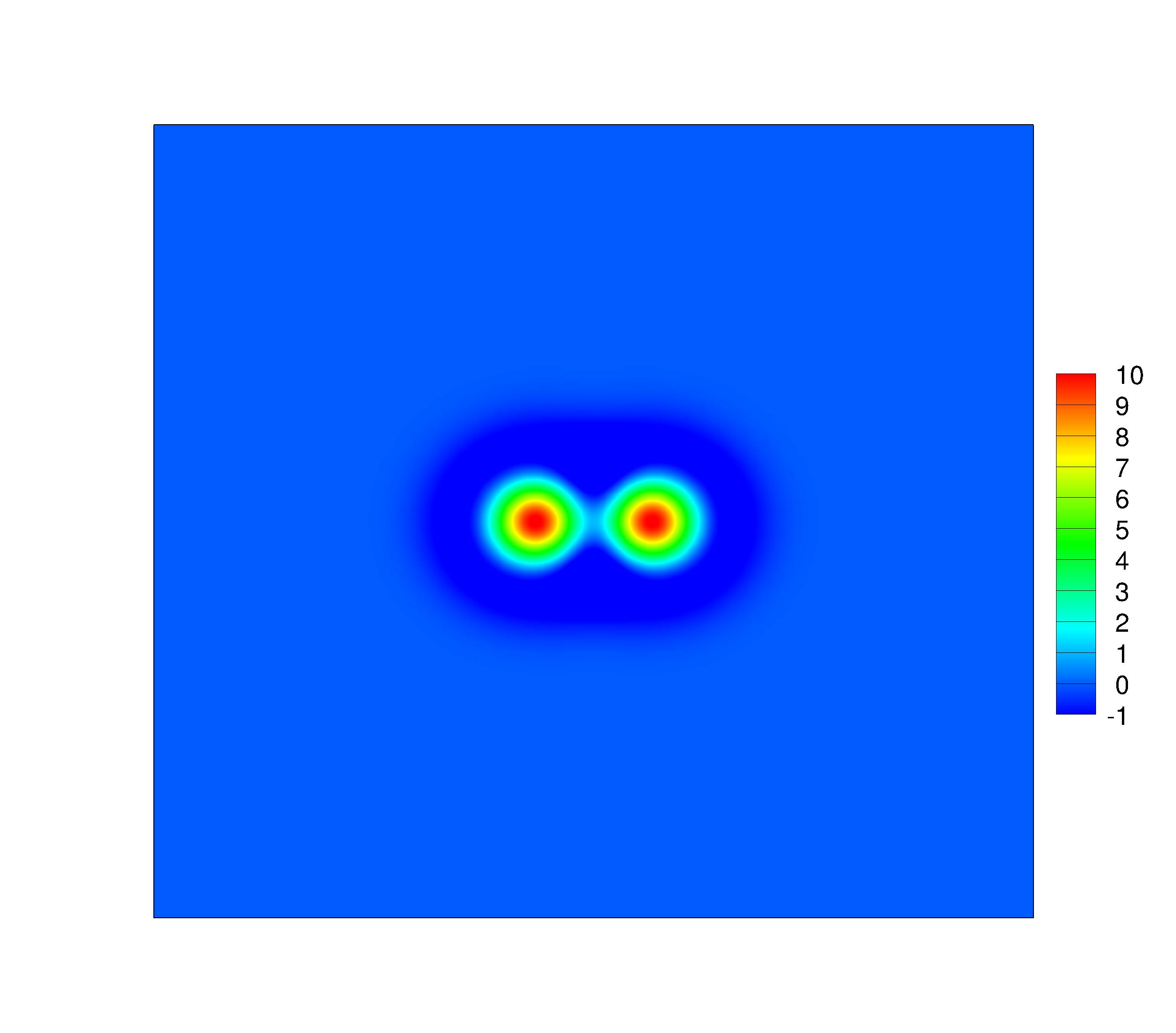}
\caption{}
\label{Fig:Flat-Taylor-vortex-a}
\end{subfigure} %
\begin{subfigure}[b]{0.4\textwidth}
\includegraphics[width=\textwidth]{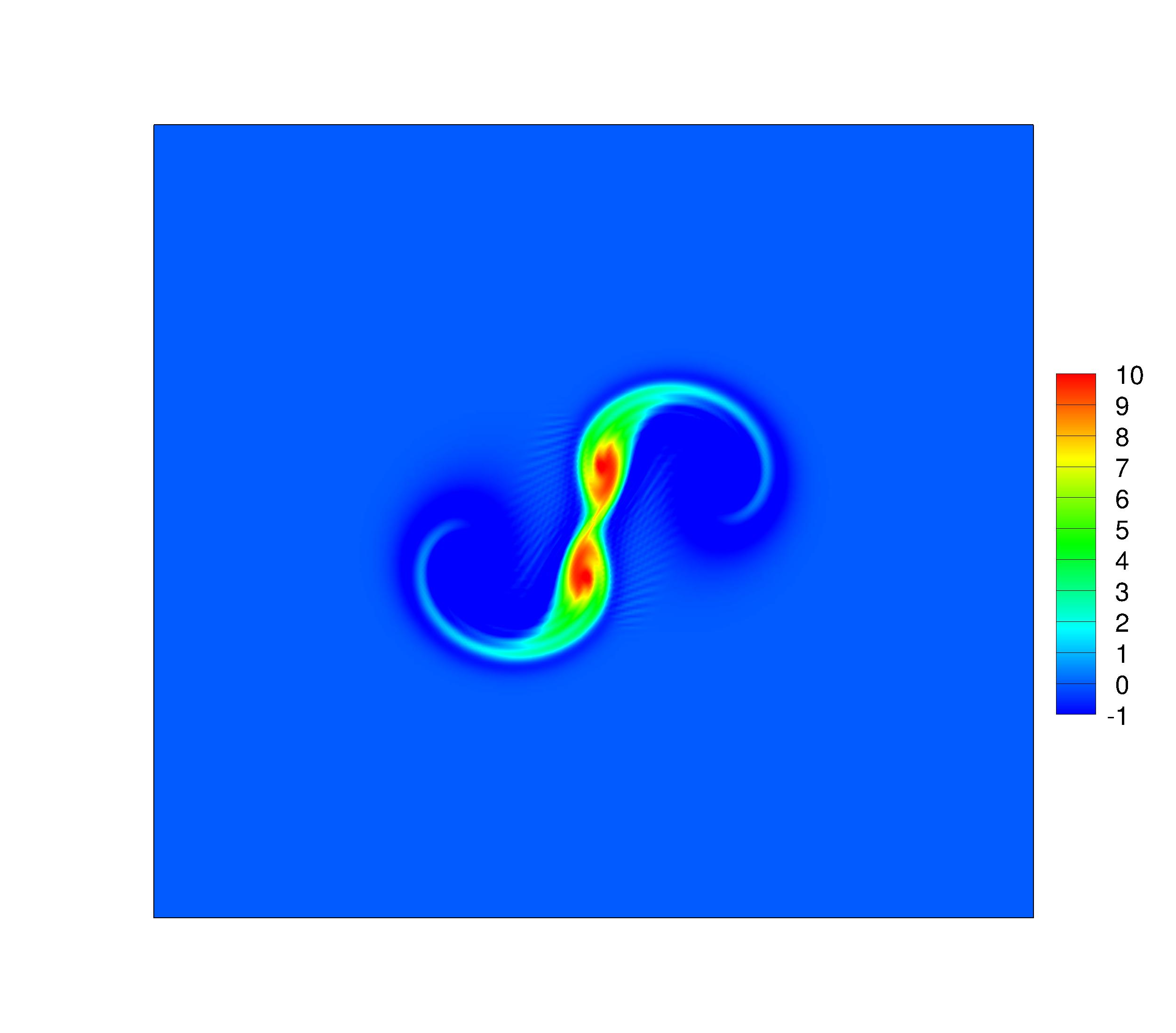} 
\caption{}
\label{Fig:Flat-Taylor-vortex-b}
\end{subfigure} %
\begin{subfigure}[b]{0.4\textwidth}
\includegraphics[width=\textwidth]{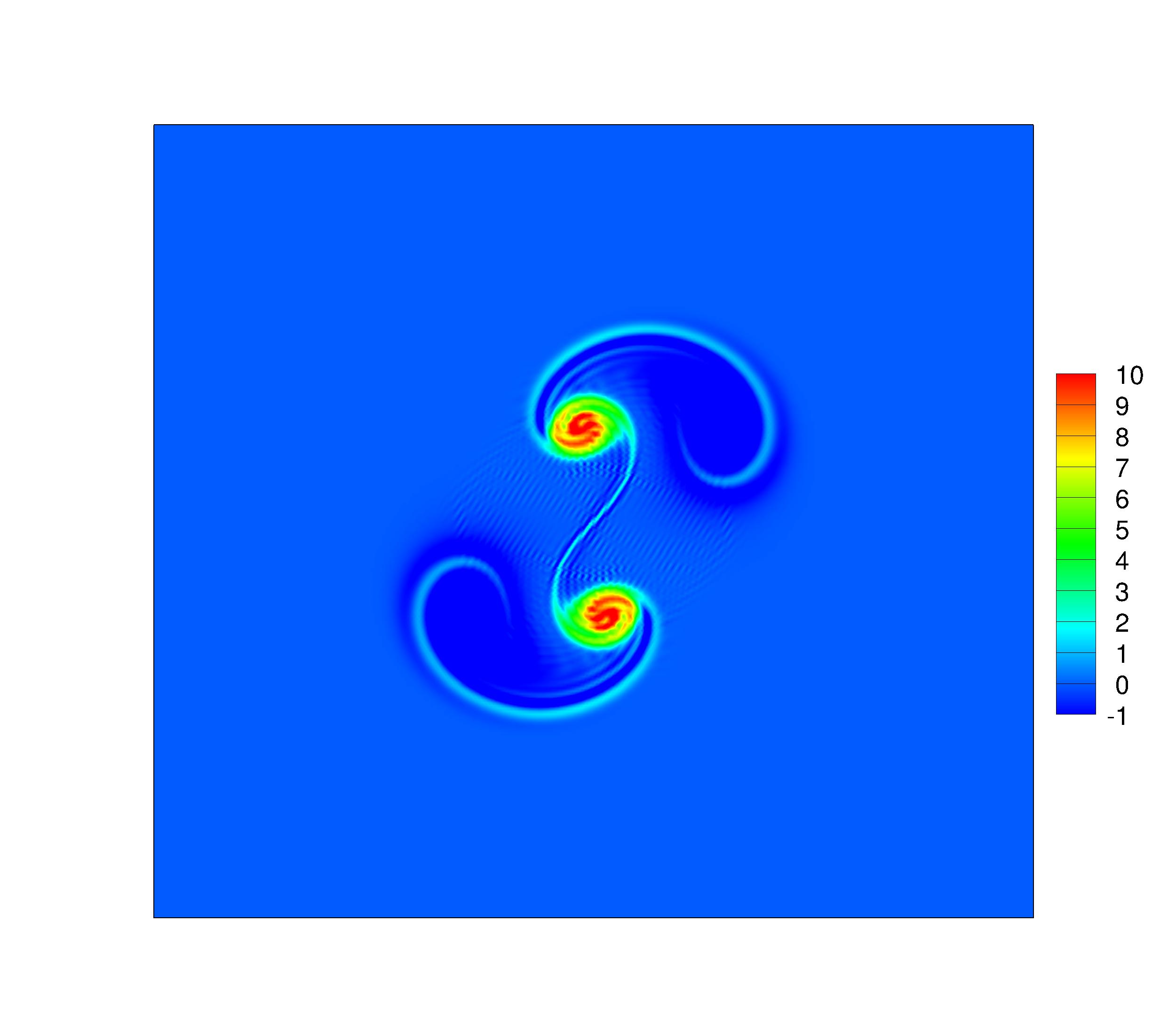}
\caption{}
\label{Fig:Flat-Taylor-vortex-c}
\end{subfigure} %
\begin{subfigure}[b]{0.4\textwidth}
\includegraphics[width=\textwidth]{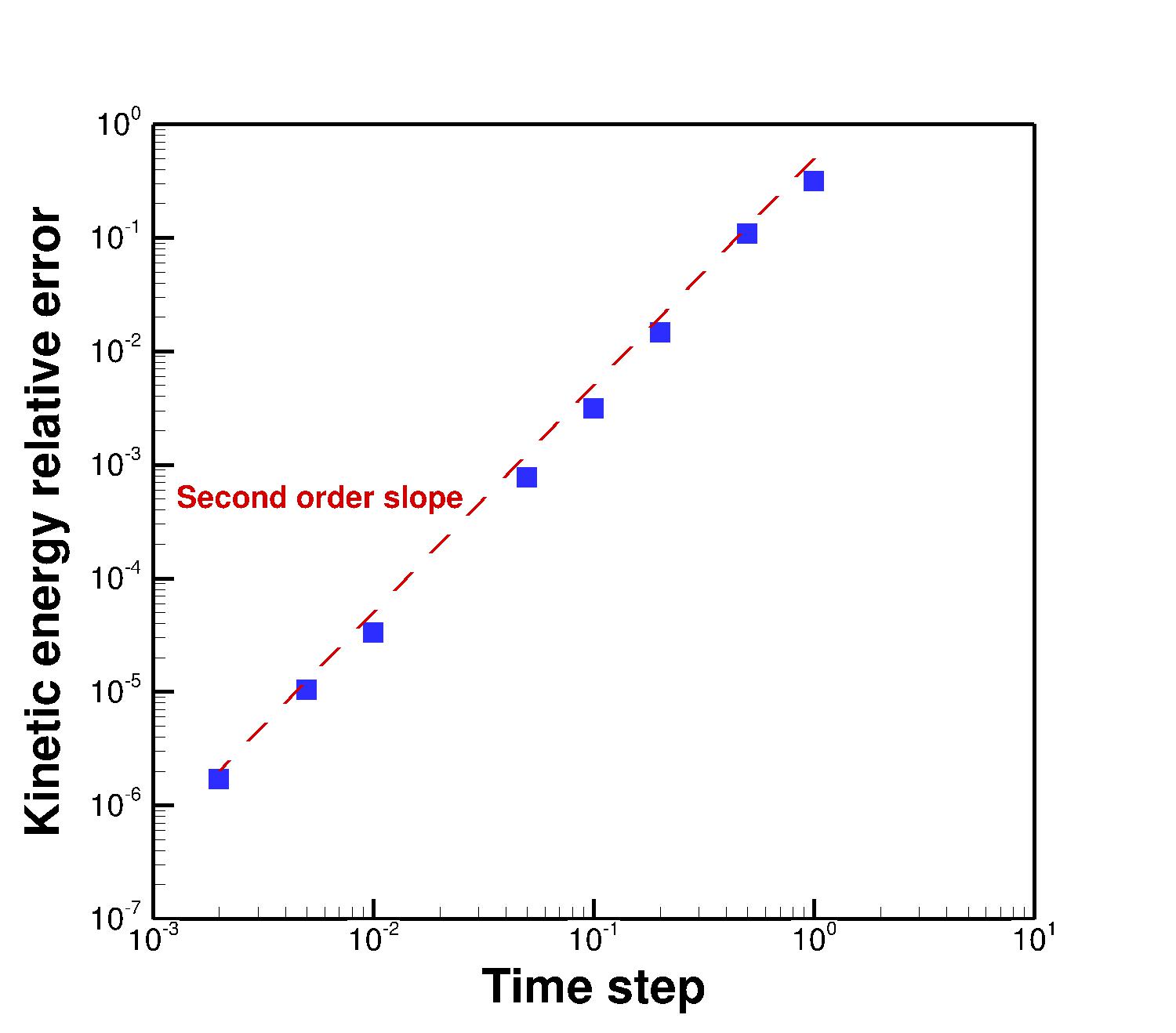} 
\caption{}
\label{Fig:Flat-Taylor-vortex-d}
\end{subfigure} %
\begin{subfigure}[b]{0.4\textwidth}
\includegraphics[width=\textwidth]{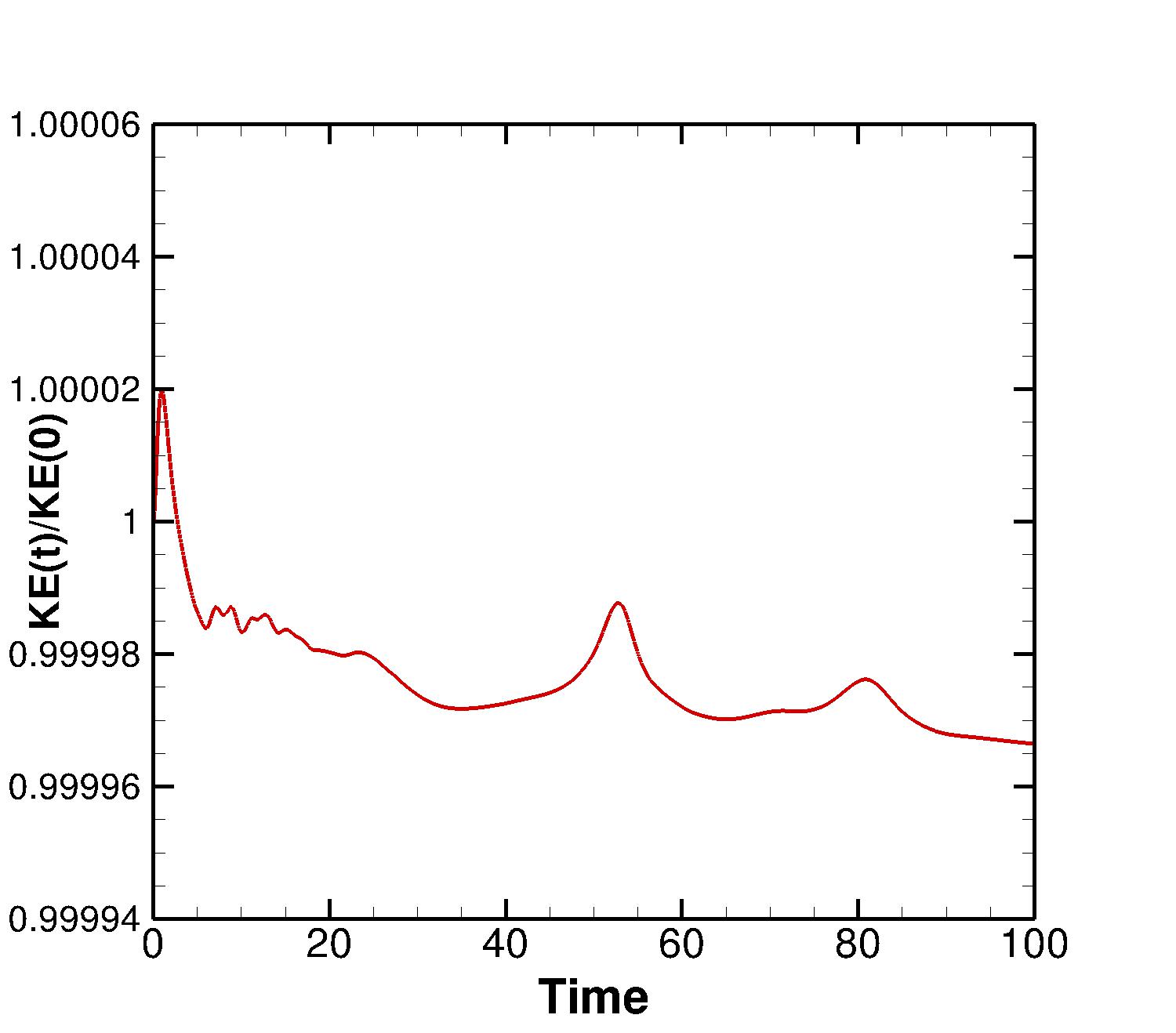} 
\caption{}
\label{Fig:Flat-Taylor-vortex-e}
\end{subfigure} %
\caption{The vorticity contour plot for two Taylor vortices using a mesh consisting of 132204 elements and a time step $\Delta t = 0.005$ at time: (\subref{Fig:Flat-Taylor-vortex-a}) T=0.0, (\subref{Fig:Flat-Taylor-vortex-b}) T=3.0 and (\subref{Fig:Flat-Taylor-vortex-c}) T=5.0. (\subref{Fig:Flat-Taylor-vortex-d}) The convergence of the relative kinetic energy error with the time step at simulation time $T=20.0$. The dashed red line represents the second order slope. (\subref{Fig:Flat-Taylor-vortex-e}) The evolution of the relative kinetic energy with time for two Taylor vortices on a hexagonal domain.}
\label{Fig:Flat-Taylor-vortex}
\end{figure}

The simulations are carried out on a mesh consisting of 132204 equilateral triangles, using various time steps in the range $[1.0-0.002]$. Fig. \ref{Fig:Flat-Taylor-vortex} shows the vorticity contour plot evolution with time, using a time step of $0.005$. The two vortices initially approach and turn over each other. The vortices then move apart, as expected, with a thin vortex sheet connecting them that disappears at longer simulation time.

The relative kinetic energy error is calculated at simulation time $T=20.0$ and is plotted versus the time step in Fig. \ref{Fig:Flat-Taylor-vortex-d}. The figure shows a second order convergence of the relative kinetic energy error over the entire range of time steps. Such time convergence rate can be due to the two-step time stepping scheme implemented during the current discretization. For practical time steps that can resolve the physics of the considered problem (e.g. $\Delta t < 0.01$), trivial relative kinetic energy error, below $0.01\%$, is observed.

Another simulation is carried out for two Taylor vortices on a hexagonal domain, aiming to compare with the Eulerian formulation developed by Pavlov et al. \cite{pavlov2011structure}. The hexagonal domain has a side length of $\pi$ and is meshed with 55296 equilateral triangular elements. The initial condition for the two Taylor vortices is the same as in Eq. \eqref{eq:taylor-vorticity-distribution} (with $G=1.0$ and $a=0.3$), and periodic boundary conditions are applied on all domain boundaries. Since the authors in \cite{pavlov2011structure} did not state the time step they used, we choose to use a time step of $0.01$ during this test case. It is expected that the time step used in \cite{pavlov2011structure} will not be significantly larger than $0.01$ since a simulation with a time step of $0.02$ was tested, however the time step was not small enough to capture the expected behavior of the two vortices, and the two vortices merged together when using the $0.02$ time step. Fig. \ref{Fig:Flat-Taylor-vortex-e} shows the change in the relative kinetic energy for extended simulation time up to $T=100$. It is observed that the kinetic energy error is within the range of $0.003\%$, which is almost three orders of magnitude smaller than the error of $2.0\%$ reported in Pavlov et al. \cite{pavlov2011structure}. 

It is worth noting that the initial increase in the kinetic energy in Fig. \ref{Fig:Flat-Taylor-vortex-e} was also observed for the previous test cases of Taylor vortices on a square domain, but was comparable to the subsequent decrease in the kinetic energy only for the time steps $\leq 0.01$. Therefore, for these time steps $\leq 0.01$, the relative kinetic energy error ($\frac{KE(0) - KE(T)}{KE(0)}$) in Fig. \ref{Fig:Flat-Taylor-vortex-d} was calculated by comparing with the kinetic energy peak at time $\sim 1.0$ instead of the initial kinetic energy $KE(0)$. This should better represent how much kinetic energy loss the implemented scheme allows.                


\subsection{Taylor vortices on a spherical surface}
\label{subsec:spherical-Taylor-vortices}

The ability of the current discretization to simulate flows over curved surfaces is explored for an inviscid flow test case over a spherical surface. The spherical surface, with radius equal to $1.0$, is approximated by a simplicial mesh consisting of flat triangles connecting the primal nodes, where the primal nodes are positioned on the smooth spherical surface (at radius = $1.0$). For each primal edge, its dual is the kinked line connecting the circumcenters of the two neighboring triangles through the primal edge's midpoint. During the discrete Hodge star operator calculations, the length of a dual edge is its length as a kinked line. On the other hand, for each primal node, its dual is the non-planar polygon consisting of sector contributions from all the flat triangles sharing the primal node, and its area is calculated accordingly as a non-planar surface area.

The test case investigated in this section is for two Taylor vortices initially positioned on a spherical surface and their evolution with time is simulated. The spherical surface domain is initialized with two vortices each have the distribution given in Eq. \eqref{eq:taylor-vorticity-distribution}, with $G=0.5$, $a=0.1$. When calculating the vorticity at any mesh node via Eq. \eqref{eq:taylor-vorticity-distribution}, the distance $r$ is measured along the sphere surface; i.e. geographical distance. The centers of the two vortices are separated by a distance of $0.4$. The simulation is carried out using a mesh containing 327680 triangular elements.

In order to recover the velocity 1-forms from the vorticity distribution, the Poisson equation \eqref{eq:taylor-vorticity-Poisson} is solved. During the Poisson equation solution, the stream function at one primal node need to be fixed in order to obtain a unique solution. Using the resulting velocity 1-forms as an initial condition, the evolution of the two vortices is then simulated using various time steps in the range $[1.0-0.05]$.

\begin{figure}
\begin{subfigure}[b]{0.5\textwidth}
\centering
\includegraphics[width=\textwidth]{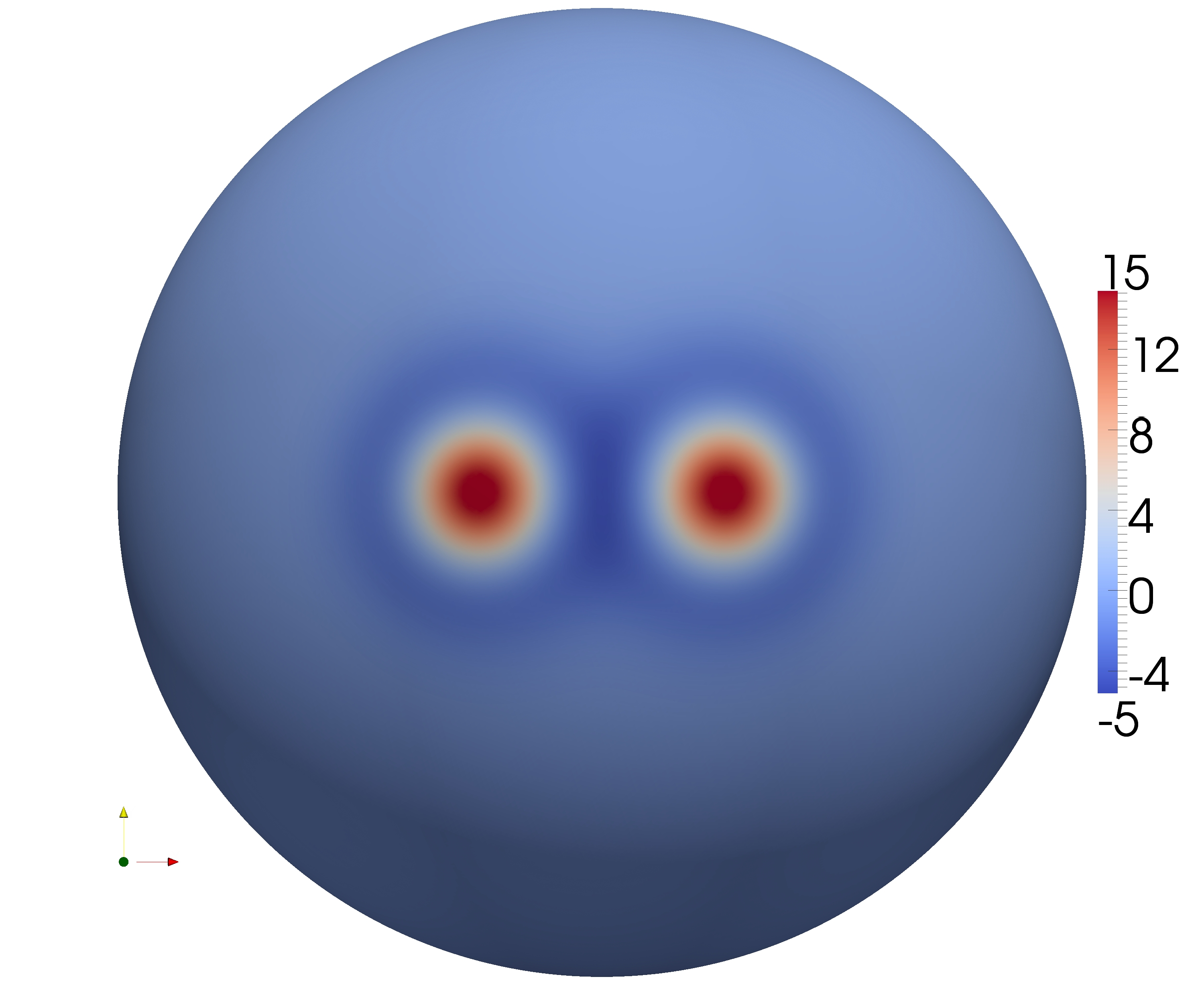}
\caption{}
\label{Fig:Sphere-Taylor-vortex-a}
\end{subfigure} %
\begin{subfigure}[b]{0.5\textwidth}
\centering
\includegraphics[width=\textwidth]{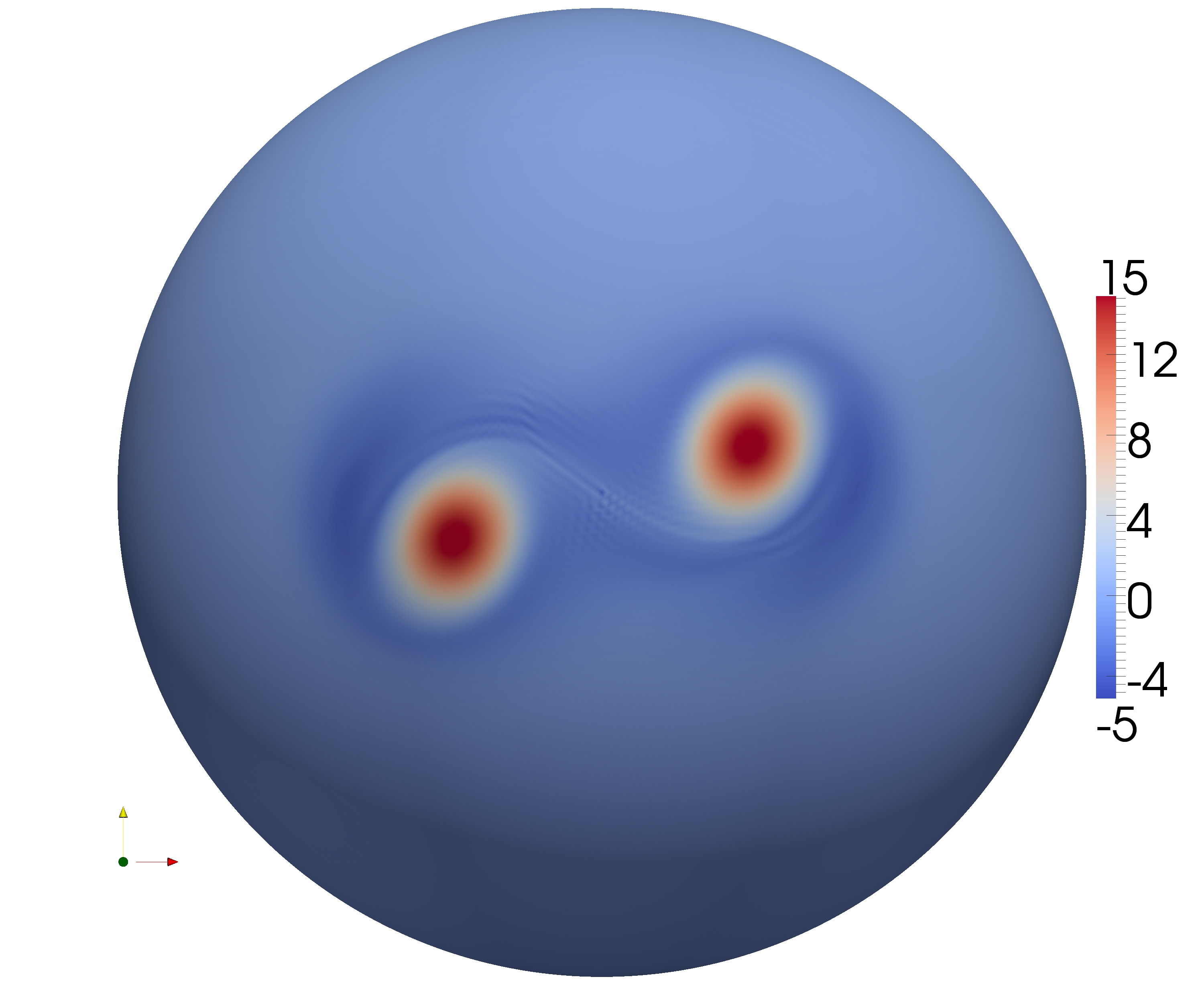} 
\caption{}
\label{Fig:Sphere-Taylor-vortex-b}
\end{subfigure} %
\begin{subfigure}[b]{0.5\textwidth}
\centering
\includegraphics[width=\textwidth]{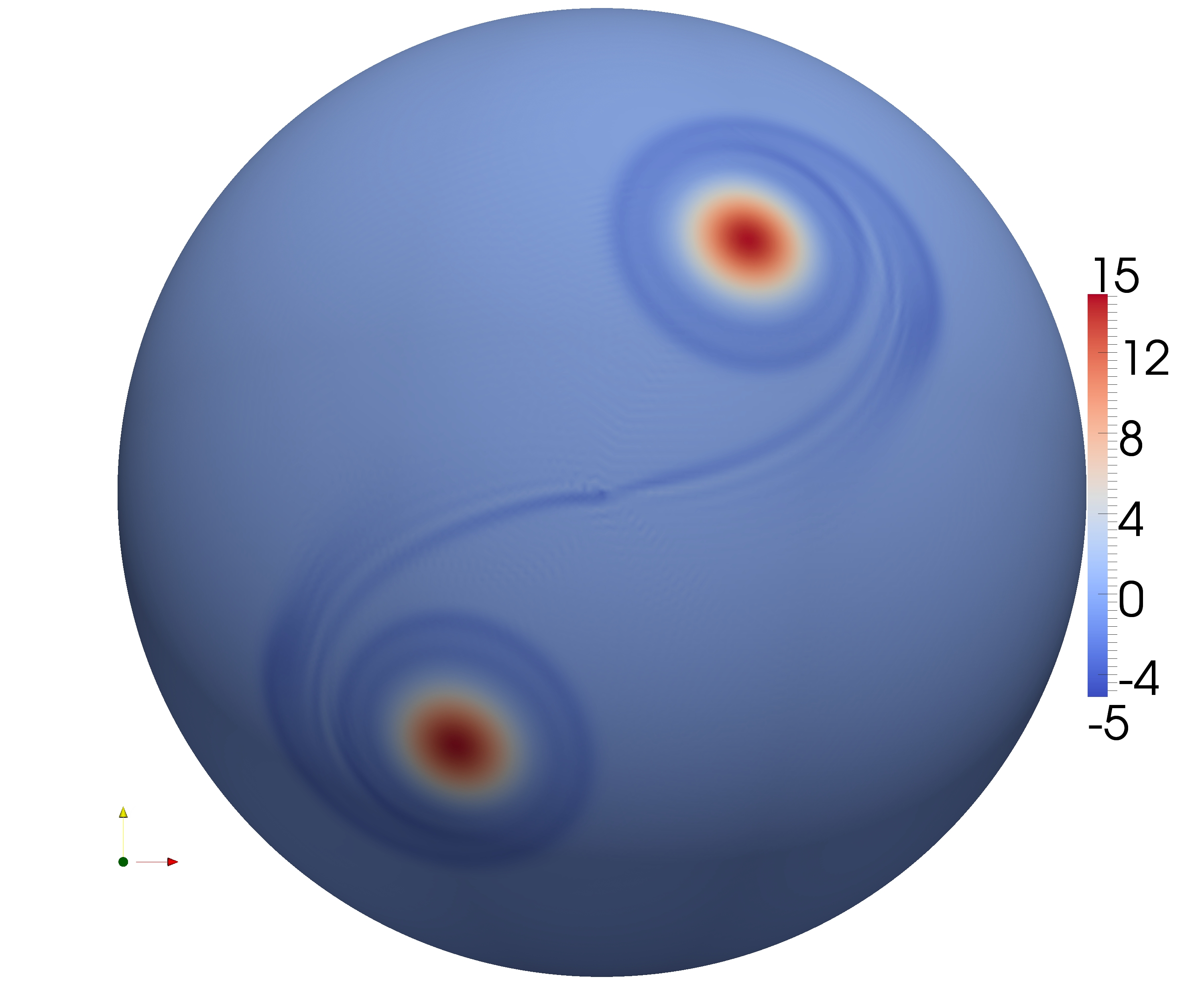}
\caption{}
\label{Fig:Sphere-Taylor-vortex-c}
\end{subfigure} %
\begin{subfigure}[b]{0.5\textwidth}
\centering
\includegraphics[width=\textwidth]{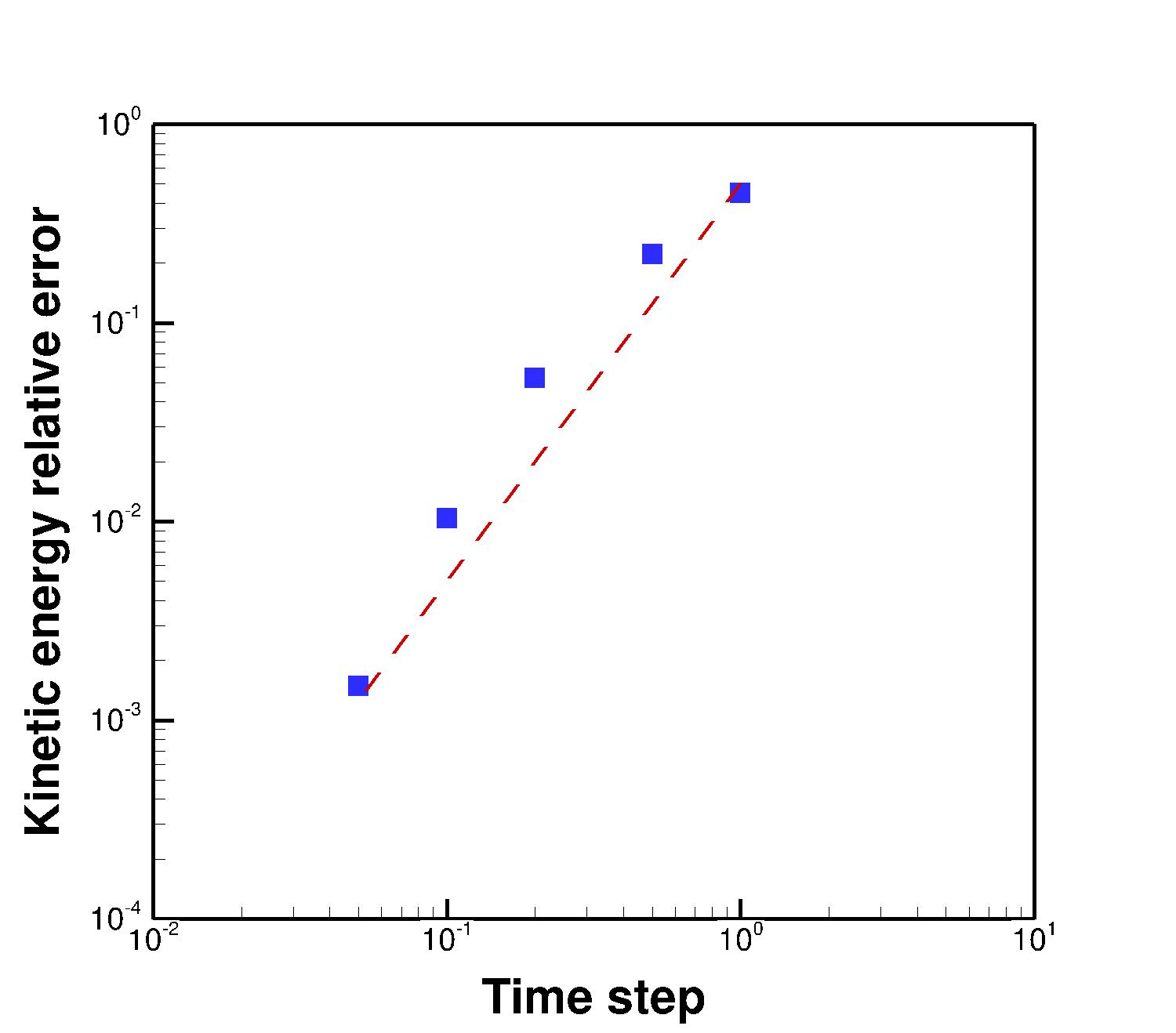}
\caption{}
\label{Fig:Sphere-Taylor-vortex-d}
\end{subfigure} %
\caption{The vorticity contour plot for two Taylor vortices on a spherical surface meshed with 327680 elements at time: (\subref{Fig:Sphere-Taylor-vortex-a}) T=0.0, (\subref{Fig:Sphere-Taylor-vortex-b}) T=2.0 and (\subref{Fig:Sphere-Taylor-vortex-c}) T=10.0. (\subref{Fig:Sphere-Taylor-vortex-d}) The convergence of the relative kinetic energy error with the time step at simulation time $T=10.0$. The dashed red line represents the second order slope.}
\label{Fig:Sphere-Taylor-vortex}
\end{figure}

Fig. \ref{Fig:Sphere-Taylor-vortex} shows the evolution of the vorticity contour plot with time. Again, the two vortices move apart with a thin vortex sheet connecting them. The convergence of the kinetic energy relative error with the time step is investigated after simulation time $T=10.0$, as shown in figure \ref{Fig:Sphere-Taylor-vortex-d}. Similar to the flow over a flat surface, a second order rate, on average, is observed for the convergence of the kinetic energy relative error with the time step. This again can be due to the two-step time stepping scheme implemented during the current discretization.     

\subsection{A ring of vortices on a spherical surface}
\label{subsec:spherical-vortices-ring}

The behavior of a ring of $N$ equidistant point vortices, having the same strength, positioned on a circle with fixed latitude on a spherical surface was previously investigated theoretically \cite{polvani1993wave}. It was shown that with such a configuration, the vortices will rotate around the z-axis in a stable fashion given that the circle's latitude $\theta$ is below a critical value and the number of vortices $N \leq 7$. For $N = 6$, the critical polar angle $\theta_c \sim 0.464$ \cite{vankerschaver2014novel}. The behavior of such a ring of vortices is simulated, where the point vortices are replaced with vortices having the distribution
\begin{equation}
\omega = \frac{\tau}{\cosh^2(\frac{3r}{a})} 
\label{Eq:vortices-ring} 
\end{equation}
with $\tau$ to be the vortex strength, $a$ is the vortex radius, and $r$ is the distance between any field point and the vortex center.

Six identical vortices, having a strength $\tau=3$ and a radius $a=0.15$, are placed on a unit sphere at latitude $\theta = 0.4$. In order to satisfy the condition that the integration of the vorticity over a spherical surface is zero, an additional vortex, with a strength $\tau = -18$ and a radius $a=0.15$, is placed at the south pole ($\theta = \pi$). The spherical surface is meshed with 81920 elements, and the simulation is conducted for an inviscid flow with a time step $\Delta t = 0.005$.

\begin{figure}
\begin{subfigure}[b]{0.5\textwidth}
\centering
\includegraphics[width=\textwidth]{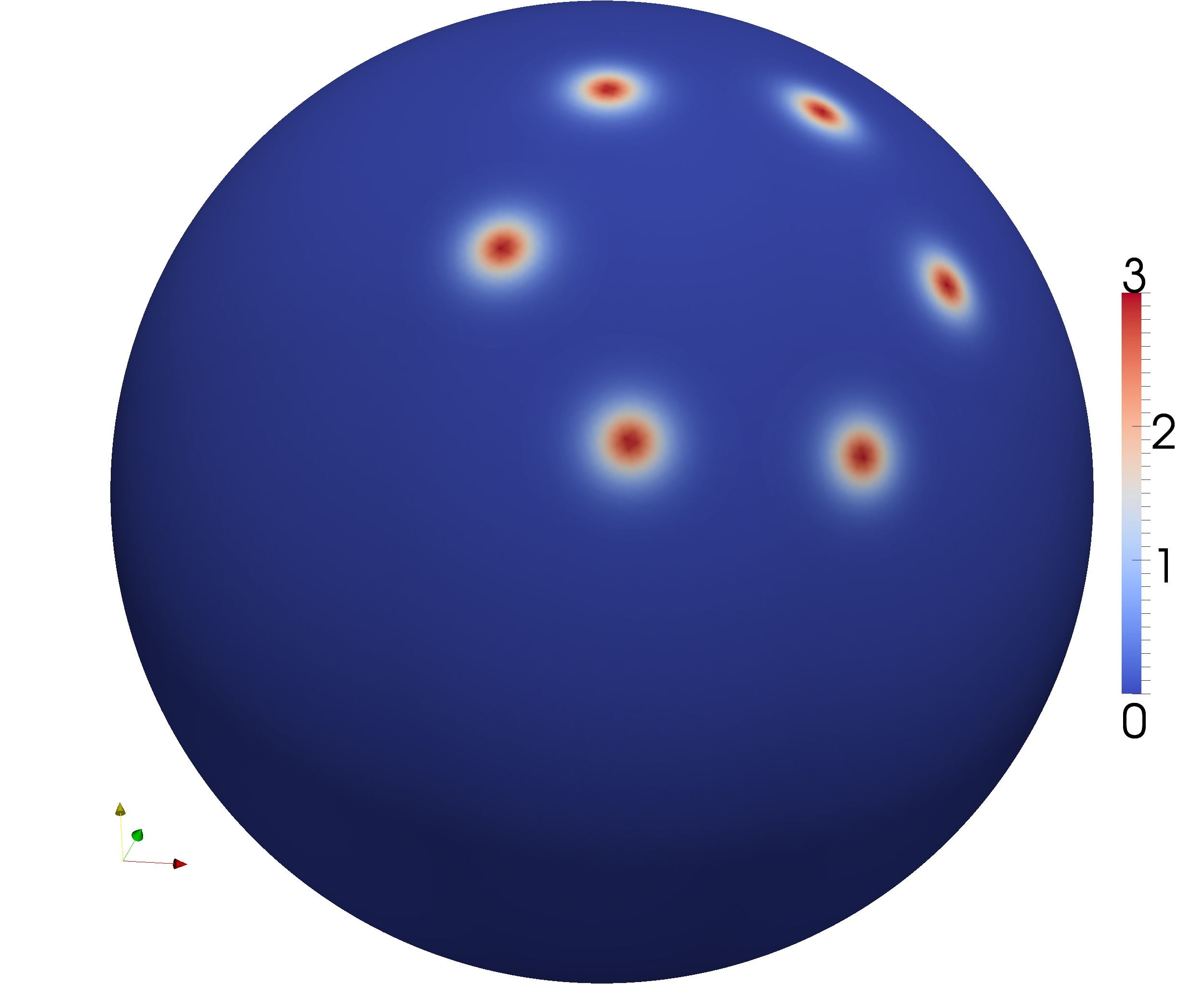}
\caption{}
\label{Fig:spherical-vortices-ring-a}
\end{subfigure} %
\begin{subfigure}[b]{0.5\textwidth}
\centering
\includegraphics[width=\textwidth]{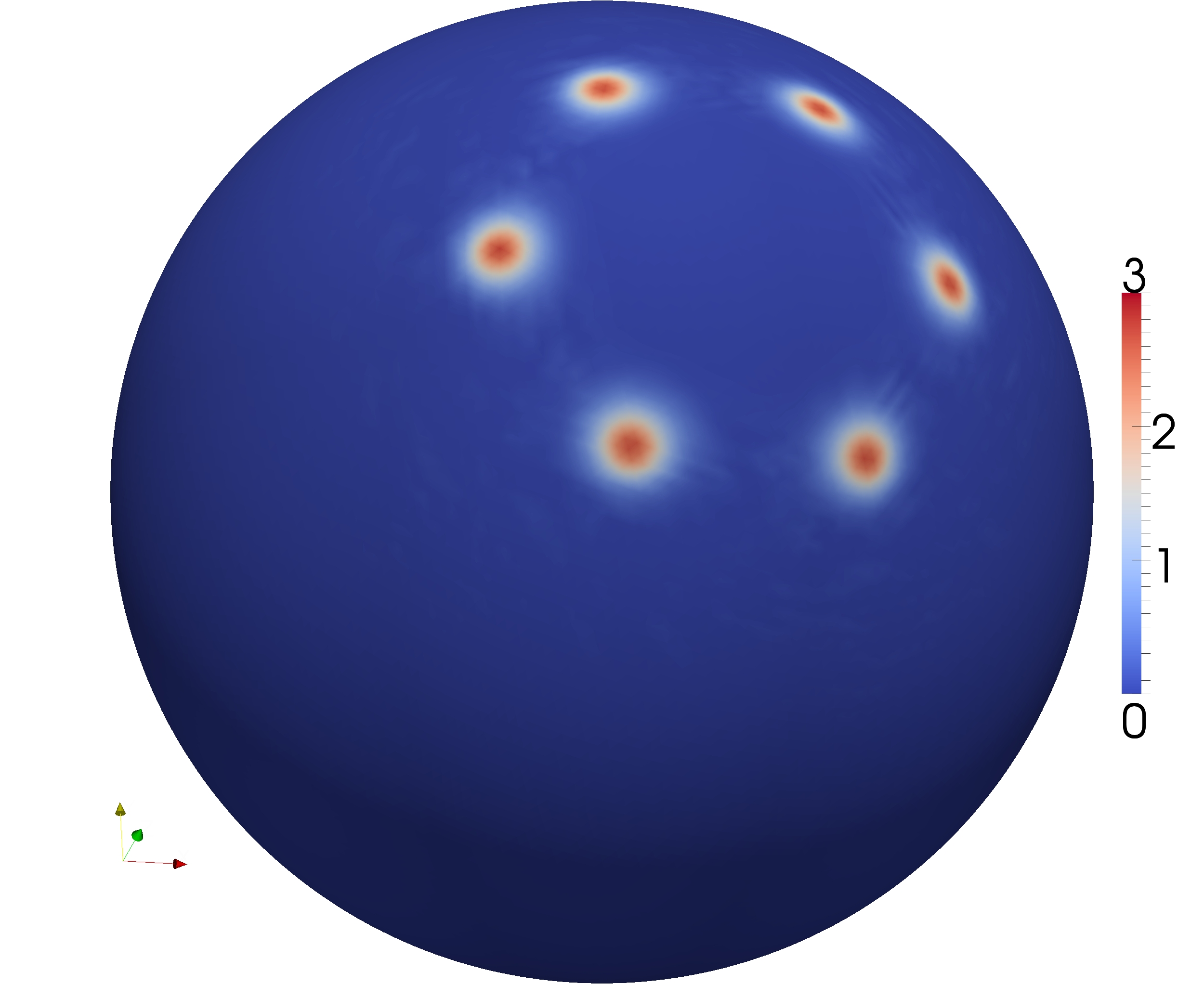} 
\caption{}
\label{Fig:spherical-vortices-ring-b}
\end{subfigure} %
\begin{subfigure}[b]{0.5\textwidth}
\centering
\includegraphics[width=\textwidth]{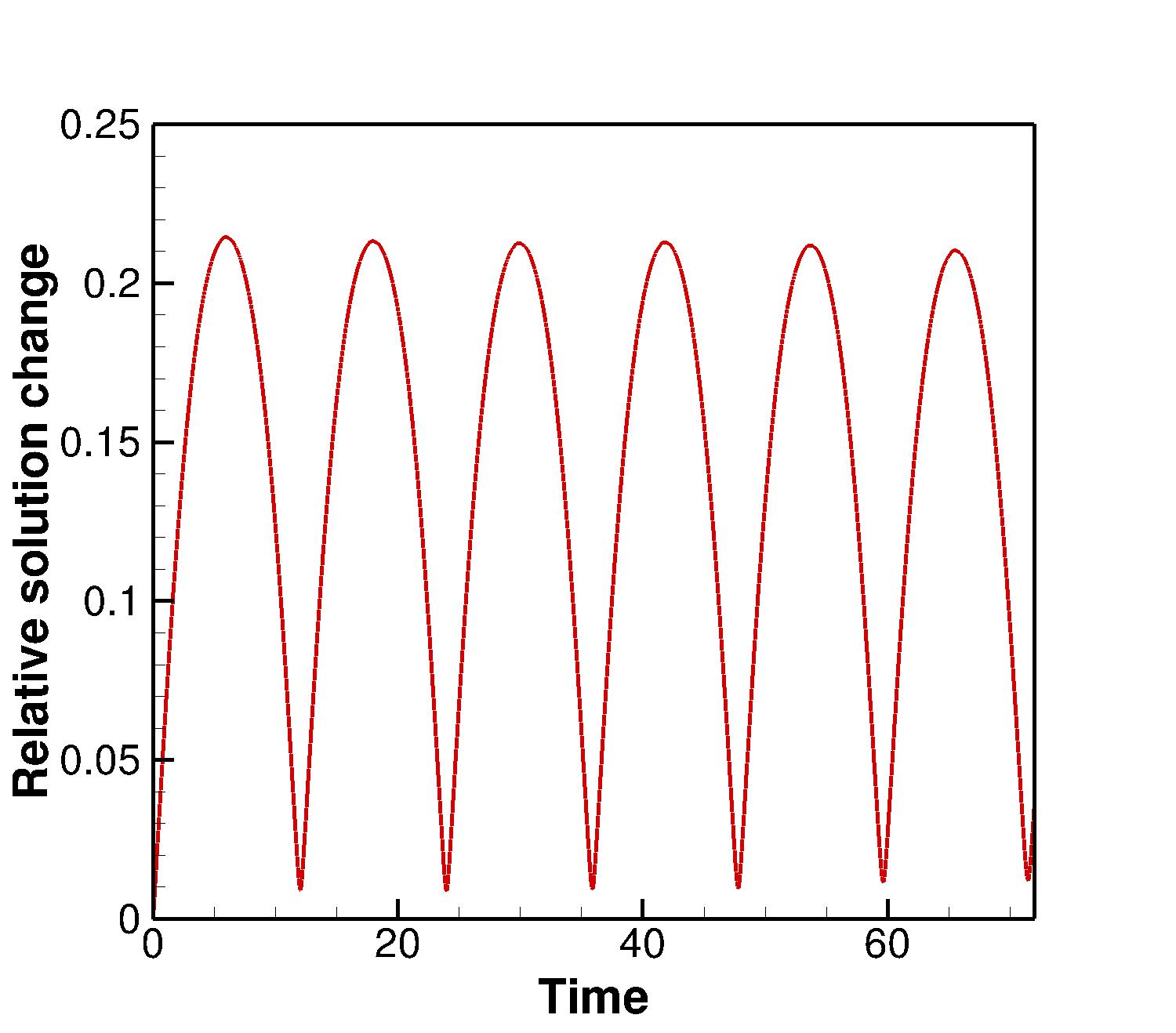}
\caption{}
\label{Fig:spherical-vortices-ring-c}
\end{subfigure} %
\begin{subfigure}[b]{0.5\textwidth}
\centering
\includegraphics[width=\textwidth]{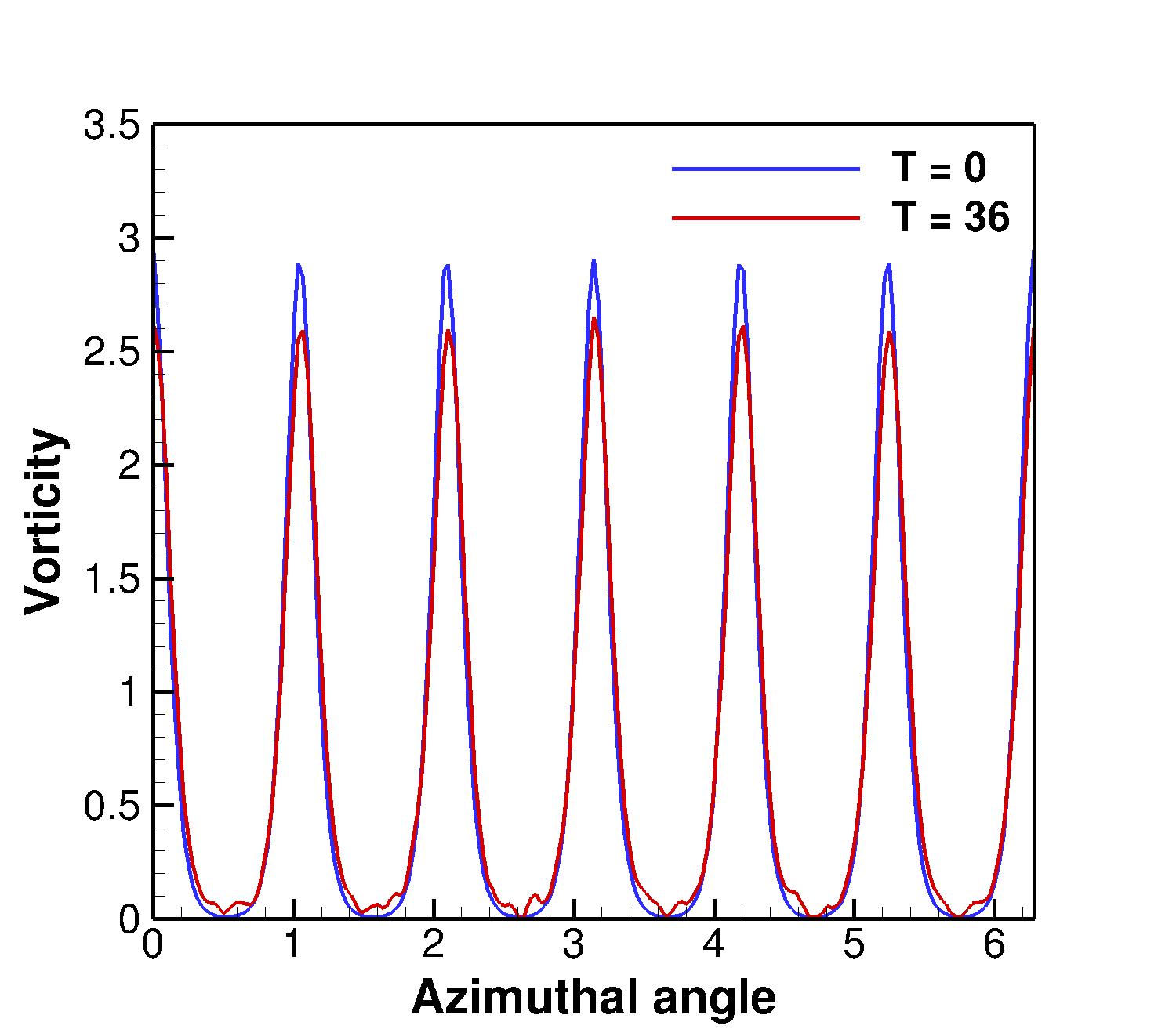}
\caption{}
\label{Fig:spherical-vortices-ring-d}
\end{subfigure} %
\caption{The vorticity contour plot for 6 vortices on a spherical surface at latitude $\theta = 0.4$ at time: (\subref{Fig:spherical-vortices-ring-a}) T=0.0 and (\subref{Fig:spherical-vortices-ring-b}) T=36.0.  (\subref{Fig:spherical-vortices-ring-c}) The relative solution change ($\frac{||U(t)-U(0)||}{||U(0)||}$) versus the simulation time. (\subref{Fig:spherical-vortices-ring-d}) The vorticity strength along a circle at latitude $\theta = 0.4$.}
\label{Fig:spherical-vortices-ring}
\end{figure}

Figs. \ref{Fig:spherical-vortices-ring-a} and \ref{Fig:spherical-vortices-ring-b} show the vorticity contour plots at time $T=0$ and $T=36$, respectively. It is observed that the vortices positions seem unchanged after such simulation time, with some flow fluctuations around the vortices due to the inviscid nature of the flow. The cyclic rotation of the vortices around the z-axis can be detected by monitoring the relative solution change, with respect to the original solution, with time. Recalling that the vector $U(t)$ contains the fluxes over all mesh edges at time $t$, the relative solution change is then defined as $\frac{||U(t)-U(0)||}{||U(0)||}$. Such relative solution change should be equal to zero each time the six vortices rotate by an angle $\pi/3$ around the z-axis. The relative solution change versus time is shown in Fig. \ref{Fig:spherical-vortices-ring-c}, which reveals the periodic nature of the vortices motion. The six vortices perform a $\pi/3$ rotation around the z-axis in a time period of almost $12$ time units. Accordingly, at time $T=36$, the six vortices have rotated by an angle $\pi$ around the z-axis. A small non-vanishing relative solution change, of almost $0.01$, is observed after each cycle, which is due to the developing flow fluctuations around the vortices, as was shown in Fig. \ref{Fig:spherical-vortices-ring-b}. Finally, the vorticity strength along the circle with latitude $\theta = 0.4$ is shown in Fig. \ref{Fig:spherical-vortices-ring-d} at simulation times $T=0$ and $T=36$. The figure indicates more quantitatively that at time $T=36$ the vortices positions are similar to the original positions due to their $\pi$ rotation around the z-axis.  The vorticity strength drop at the center of all the six vortices is due to the developed flow fluctuations around the vortices. Recalling that the vorticity integration over the spherical surface is always by construction maintained at zero, and noting that the strength of the single vortex at the south pole only changed by $0.002 \%$ at time $T = 36$, the vorticity developed around the six vortices due to flow fluctuations is compensated from the six vortices themselves. In regards to the kinetic energy, the relative change in the kinetic energy at time $T=36$ is $\frac{KE(T=0) - KE(T=36)}{KE(T=0)} = 9.0 \times 10^{-6}$.                     

\section{Conclusions}
\label{sec:Conclusions}

A conservative discrete exterior calculus discretization of Navier-Stokes equations was developed. The Navier-Stokes equations were first rewritten using the exterior calculus notation. The expression of Navier-Stokes equations using the exterior calculus notation was derived from the standard vector calculus formulation and verified against the coordinate invariant form in terms the exterior and Lie derivatives.

The discretization was carried out through the substitution by the discrete exterior calculus operators defined on simplicial meshes. Both 2D and 3D discretizations were developed, but with numerical implementations only for flows over surfaces. The main distinction between the developed discretization and previous unstructured conservative discretizations was in the convective term. In the 2D case, the current convective term discretization is different from all previous DEC-based discretizations and most of the covolume method discretizations. The developed discretization has similarities with the covolume discretization by Perot \cite{perot2000conservation} only for the special case of a structured-triangular mesh. Nevertheless, unlike all covolume discretizations, the presented discretization is capable of simulating flows over both flat and curved surfaces. In regards to the 3D case, the current convective term discretization is different from all previous unstructured conservative discretizations. An additional merit of the presented methodology is the manipulation of the convective term through algebraic discretization of the interior product operator and a combinatorial discretization of the wedge product. Such approach paves the way to explore the application of the finite element exterior calculus method to discretize Navier-Stokes equations. Moreover, it gives insight into the discretization of similar convective terms in other physics problems; e.g. the magnetohydrodynamics governing equations. 

Several 2D simulation experiments were carried out to benchmark the discretization performance. The convergence of the velocity 1-forms (i.e. fluxes) was found to be of second order for structured-triangular meshes and of first order otherwise. This is in agreement with previous theoretical estimations developed for the covolume method. In regards to the conservation properties, due to the discretization construction, both the mass and the vorticity are conserved locally and globally up to the machine precision. The kinetic energy relative error converged in a second order fashion with the mesh size for flows over flat surfaces. The convergence of the kinetic energy relative error with the time step was also found to be of a second order for flows over both flat and curved surfaces. Such conservation properties, the ability to simulate flows over both flat and curved surfaces and the relatively small size of the linear system make the presented discretization attractive for both physics and engineering applications.

\section*{Acknowledgments}

This research was supported by the KAUST Office of Competitive Research Funds under Award No. URF/1/1401-01.

\appendix

\section{The complementary contribution to the dual 2-cells boundary operator}
\label{app:boundary-operator}

After defining the velocity 1-forms on primal/dual mesh entities and substituting with the appropriate discrete operator, the discretized Navier-Stokes equation was expressed, as in Eq. \eqref{eq:discreteNS01}, as  

\begin{multline}
\label{eq:discreteNS01-app}
\underline{[-\textrm{d}^T_0]} \frac{U^{n+1} - U^n}{\Delta t} - \mu \underline{[-\textrm{d}^T_0]} \ast_1 \textrm{d}_0 \ast^{-1}_0 \left[[-\textrm{d}^T_0] U + \textrm{d}_b V \right] \\ +  \underline{[-\textrm{d}^T_0]} \ast_1 W_v \ast^{-1}_0 \left[[-\textrm{d}^T_0] U + \textrm{d}_b V\right] =0, 
\end{multline}

For the dual 2-cells touching the domain boundary; e.g. the 2-cell whose dual is the primal nodes 0, 1, 3, 4, .. in Fig. \ref{fig:mesh2D}, the discrete operators $[-\textrm{d}^T_0]$ in the viscous and convective terms are complemented by the operator $\textrm{d}_b$ that closes such dual 2-cells boundaries by the primal boundary edges. Such domain boundary contribution vanishes, however, for other $[-\textrm{d}^T_0]$ operators, underlined in the above equation. Regarding the time derivative term, since the entries of the tangential velocity forms vector $V$ are calculated at an intermediate time step, the domain boundary contributions complementing $[-\textrm{d}^T_0] U^{n+1}$ and $[-\textrm{d}^T_0] U^n$ will then cancel each other.

In regards to the viscous term, starting from its smooth exterior calculus form; i.e. $\mu \textrm{d} \ast \textrm{d} \ast \textrm{d} \mathbf{u}^\flat$, it can be expressed as $\mu \textrm{d} \alpha$, with $\alpha =  \ast \textrm{d} \ast \textrm{d}  \mathbf{u}^\flat$. Considering the domain boundary contribution, the discrete viscous term is then expressed as $\mu \left[[-\textrm{d}^T_0] A + \textrm{d}_b A'\right]$, with $A$ to be the vector containing the discrete $\alpha$ 1-forms defined on the dual edges, and $A'$ as the vector containing the discrete 1-forms $\alpha'$. Similar to $\alpha$, the smooth form $\alpha'$ is defined as $\alpha' = \ast \textrm{d} \ast \textrm{d} \mathbf{u}^\flat$, whereas its discrete version is defined however on the primal edges. It follows accordingly, based on the diagram in Eq. \eqref{eq:cochncmplx2d}, that the discretization of $\textrm{d} \mathbf{u}^\flat$ included in the $\alpha'$ form is defined on the primal triangles. Since the smooth velocity form $\mathbf{u}^\flat$ is retrieved through Whitney map interpolation as a constant form over each triangle, the exterior derivative of such constant velocity form then vanishes; i.e. $\int_{\sigma_2} \textrm{d} \mathbf{u}^\flat = 0$. This implies that $\alpha'$, and therefore the domain boundary contribution, vanishes for the viscous term. Using a similar argument for the convective term $\textrm{d} \ast (\mathbf{u}^\flat \wedge \ast \textrm{d} \mathbf{u}^\flat)$, it follows that the discretization of $\textrm{d} \mathbf{u}^\flat$, in the wedge product, is also defined on the primal triangles, and therefore is equal to zero. Accordingly, the domain boundary contribution to the first (underlined) $[-\textrm{d}^T_0]$ operator in the convective term also vanishes.

\section*{References}
\bibliographystyle{elsarticle-num} 
\bibliography{NavierStokesDEC}





\end{document}